\documentclass[a4paper,aps,showpacs,nofootinbib]{revtex4}
\usepackage{graphicx}
\usepackage{amssymb}   
\usepackage{bm}        
\usepackage{amsmath}

\def\beq{\begin{equation}}
\def\eeq{\end{equation}}

\def\beqn{\begin{eqnarray}}
\def\eeqn{\end{eqnarray}}
\def\nn{\nonumber\\}

\def\e{\vec\epsilon}

\newcommand{\etal}{{\it et al.}}
\newcommand{\Gk} {\rlap{\hspace{0.10ex}\raisebox{0.2ex}{/}} k}
\newcommand{\Ge} {\rlap{\hspace{0.0ex}/} \epsilon}

\begin{document}

\title {Incoherent pion photoproduction on the deuteron in the first
resonance region}

\author{M.~I.~Levchuk}
\affiliation {B.I. Stepanov Institute of Physics, 220072 Minsk,
Belarus}

\author{A.~Yu.~Loginov}
\author{A.~A.~Sidorov}
\author{V.~N.~Stibunov}
\affiliation {
Nuclear Physics Institute, 634050 Tomsk,
Russia
}

\author{M.~Schumacher}
\affiliation{
Zweites Physikalisches Institut, Universit\"at G\"ottingen
   D-37077 G\"ottingen, Germany}


\begin{abstract}
Incoherent pion photoproduction on the deuteron is studied in the
first resonance region. The unpolarized cross section, the beam
asymmetry, and the vector and tensor target asymmetries are
calculated in the framework of a diagrammatic approach. Pole
diagrams and one-loop diagrams with $NN$ scattering in the final
state are taken into account. An elementary operator for pion
photoproduction on the  nucleon is taken in various on-shell forms
and calculated  using the SAID  and MAID multipole analyses. Model
dependence of the obtained results is discussed in some detail. A
comparison with predictions of other works is given. Although a
reasonable description of many available experimental data on the
unpolarized total and differential cross sections and photon
asymmetry has been achieved, in some cases a significant
disagreement between the theory and experiment has been found.
Invoking known information on the reactions $\gamma d\to\pi^0 d$
and $\gamma d\to np$ we predict the total photoabsorption cross
section for deuterium.  We find that our values strongly
overestimate experimental data in the vicinity of the $\Delta$
peak.

\end{abstract}


\maketitle

\section{Introduction}
\label{intr}

Comprehensive measurements of total and differential cross
sections of inclusive, coherent, and incoherent $\pi^0$
photoproduction from the deuteron in the energy region from 140 to
792 MeV were carried out at MAMI \cite{krusche99,siodl01}. It was
found that the coherent data are in good agreement with
theoretical predictions.  However, in the case of the incoherent
cross sections the situation was found to be much less
satisfactory. The theoretical predictions from Refs.\
\cite{laget81,schmidt96} in the $\Delta$ region exceeded the
experimental data significantly.

The above mentioned disagreement may be indicative of shortcomings in
the approaches of Refs.~\cite{laget81,schmidt96}.
The model developed in Ref.\ \cite{schmidt96} seems to
be oversimplified because it takes into account the pole diagrams
only.  It is known that nucleon-nucleon final state interaction (FSI)
is extremely important in incoherent pion photoproduction especially for
small pion angles (see, e.g., Refs.\ \cite{laget81,lps96,lsw00}).  Although
FSI was incorporated in the model of Ref.\ \cite{laget81}, it
nevertheless failed to reproduce the data. A possible reason for this
might be that Laget used in his calculations of the $\gamma d\to
\pi^0 np$ process the well-known Blomqvist-Laget (BL) parametrization
\cite{BlLag77} of the pion photoproduction amplitude on the nucleon.
This parametrization gives a good fit to the amplitude of charged
pion photoproduction.  But it does not provide a satisfactory
description of $\pi^0$ production from the proton. Because data on
$\pi^0$ production from the neutron are absent there is no
possibility to check the reliability of the BL model in the
description of this channel. An
attempt to remedy this defect made in Ref.\ \cite{sabutis83}
led to a $\pi^0$ photoproduction
operator which is not very suitable for the use in nuclear
calculations.

This unsatisfactory situation has stimulated a number
of new theoretical investigations of the reaction $d(\gamma,\pi)NN$
\cite{LSW00,darwish_diss,darwish03,darwish04,darwish05_1,darwish05_2,
darwish05_3,darwish05_4,ArFix05,FixAr05}.
In Ref.~\cite{LSW00}, a computation of the differential and
total cross sections for this process in the  first resonance region
was presented.  The main difference between the approaches from
Refs.~\cite{LSW00} and \cite{laget81} was,
that a more realistic version of an elementary
pion photoproduction operator was used in the former.
It was taken in the standard CGLN form \cite{CGLN} with
four partial amplitudes $F_i$ calculated
with the use of the SAID \cite{arnd96} and MAID \cite{drechsel99}
multipole analyses. The model provided
a satisfactory description of the data from
Refs.~\cite{krusche99,siodl01}.
Unfortunately, in Ref.~\cite{LSW00} there was an error in coding
the amplitude for the charged channels so that the reasonable
description of data on the $\pi^-$ channel should be considered
as accidental.

In a series of articles by Darwish
\etal~\cite{darwish_diss,darwish03,darwish04,darwish05_1,darwish05_2,
darwish05_3,darwish05_4}, the study of incoherent pion photoproduction
in the $\Delta$-resonance region was continued.
The authors used the elementary operator
proposed in Ref.~\cite{schmidt96} which is quite similar
to the BL operator except for slight differences in parameter values.
Reasonable description of the available data on the
total and differential cross sections
was achieved in Ref.~\cite{darwish03}.
For the first time, an attempt to analyse polarization
observables in the reaction $d(\gamma,\pi)NN$ was made
in Refs.~\cite{darwish04,darwish05_1,darwish05_2,
darwish05_3,darwish05_4}. The beam asymmetry $\Sigma$ for linearly
polarized photons, target asymmetries $T_{IM}$, and beam-target
asymmetries were discussed in those papers. However, many
conclusions drawn at the analysis of the polarizations were
wrong as it was explained in full detail in Ref.~\cite{ArFix05}.

One more analysis of this process was presented in Ref.~\cite{ArFix05}
where formal expressions for observables
in incoherent pion photoproduction were given,
as well as in a subsequent article \cite{FixAr05}. Using the MAID model
for an elementary production operator, the authors studied
the inclusive reaction from threshold up to 1 GeV.
They obtained quite satisfactory agreement with the data similar to
that achieved in Refs.~\cite{darwish04,darwish05_1,darwish05_2,
darwish05_3,darwish05_4} although considerable deviation from the
predictions of the latter articles  was found for
many polarization asymmetries. A part of the deviation is a
consequence of  the use of wrong formal expressions in those articles.
But in the cases when the right expressions were used, the origin
of the deviation remained to be unclear.

It should be noted that in neither of the works
\cite{darwish03,darwish04,darwish05_1,darwish05_2,
darwish05_3,darwish05_4,FixAr05}, the question on
model dependencies of the calculations was studied.
Only the sensitivity of the predictions to the choice of
the model for $NN$ interaction was investigated and
was found to be very small. An analogous result was
also reported in Ref.~\cite{LSW00}.
But the dependencies mentioned do exist. They mainly stem
from the elementary production operator. First, two
now available multipole analyses, SAID and MAID, are
not equivalent and give different results for observables.
The size of the deviation depends on the kinematic region
and on the observable under consideration.
For example, the total cross sections produced by the SAID
SM04K solution at the total energy of 1232 MeV are
$267~\mu b$, $270~\mu b$, $212~\mu b$, and
$240~\mu b$ for the $\pi^0p$, $\pi^0n$, $\pi^+n$, and $\pi^-p$
channels, respectively. The corresponding numbers given
by the MAID03 solution are $279~\mu b$, $281~\mu b$, $215~\mu b$,
and $244~\mu b$.
It is clear that this sensitivity to the choice of
the analysis will be seen in the reaction
$d(\gamma,\pi)NN$ too.
Even different solutions for the same analysis
(e.g., MAID00 or MAID03) give different results that
leads to additional model dependence in this reaction.
Second, both the SAID and MAID models are
the parametrizations of an on-shell production operator.
The latter depends on four invariant amplitudes
\cite{CGLN}. Different options are possible for these amplitudes
that are equivalent in the on-shell case.
However, this equivalence is broken in deuteron (or more
generally, in nuclear) calculations when one nucleon
or both of them are off their mass shells.
Though it is difficult to give precise numerical account for
the off-shell effects, we should at least estimate
possible uncertainties of the results introduced by them.

We, therefore, motivate the present work by the following reasons.
First, we would like to compare our results with those from
Refs.~\cite{darwish03,darwish04,darwish05_1,darwish05_2,
darwish05_3,darwish05_4} and Ref.~\cite{FixAr05} in the
framework of an analogous approach, i.e. in the impulse
approximation with FSI effects, because any new calculation
can serve as an independent check of which of the above
models is valid. Second, we want to estimate possible
uncertainties in theoretical predictions for observables.
These uncertainties should be kept in mind when extracting
information on the amplitude of pion photoproduction
on the neutron from deuteron data.

The paper is organized as follows. In Sect.\ \ref{kinema},
the kinematic relations used for the calculations and definitions
for observables are briefly reviewed. A description of the
theoretical model and its ingredients is given in Sect.\
\ref{theory}. Section \ref{results} contains the results
on the differential and total cross sections as well as beam and target
asymmetries with a special emphasis on possible uncertainties of the results.
In the same section we also compare our predictions with
data available in the considered kinematic region and
with results of other approaches given in Refs.~
\cite{darwish03,darwish04,darwish05_1,darwish05_2,
darwish05_3,darwish05_4,FixAr05}.
Different parametrizations for
the elementary photoproduction operator are presented in
Appendix~\ref{operator}.

\section{Kinematics and definitions of observables }
\label{kinema}

Let us denote by $k=(k^0,\vec k),~p_d=(\varepsilon_d,\vec
p_d),~q=(\varepsilon_\pi,\vec q),~p_1=(\varepsilon_1,\vec p_1)$ and
$p_2=(\varepsilon_2,\vec p_2)$ the four-momenta of the initial photon
and deuteron and the final pion and nucleons, respectively.
A symbol $E_\gamma$ is reserved for the lab photon energy
($k^0_{lab}=E_\gamma$) and a symbol $\omega$ will be used for
the photon energy in the $\gamma d$ center-of-mass (c.m.)  frame
($k^0_{cm}=\omega=E_\gamma M/W_{\gamma d}$) with $W_{\gamma
d}=\sqrt{M^2+2ME_\gamma}$ and $M$ being the deuteron mass.

We take as  independent kinematic  variables the
photon energy and pion momentum $\vec q$ in the used frame of
reference (generally, the lab or c.m.  frame) and the angles
$\Theta_{\vec P}$ and $\phi_{\vec P}$ of one of the nucleons in the
c.m. frame of the final nucleon-nucleon pair.  Using the equality
\beqn
\label{P_cm}
W_{NN}=2\varepsilon_P=2\sqrt{{\vec  P}^2+m^2}=\sqrt{(k+p_d-q)^2},
\eeqn
where $m$ is the nucleon mass, one can find
the momentum $\vec P$. After boosting the momenta $\vec P$ and
$-\vec P$ with the velocity $(\vec k+\vec p_d -\vec q)/
(k^0+\varepsilon_d -\varepsilon_\pi)$ the momenta of the outgoing
nucleons are obtained and, therefore, the kinematics is totally
determined.

The differential cross section is given by
\beqn
\label{dcs4}
\frac {d\sigma }{d\vec q d\Omega_{\vec  P}}=
\frac 1{(2\pi)^5}
\frac {m^2\varepsilon_d |{\vec P}|}
      {8k\cdot p_d~\varepsilon_\pi \varepsilon_P}
~~\frac 16
\sum _{m_2m_1\lambda m_d}
| \langle m_2m_1| T| \lambda m_d \rangle |^2,
\eeqn
where $m_2$, $m_1$, $\lambda$, and $m_d$ are spin states of
the two nucleons, photon, and deuteron, respectively.
To obtain the inclusive differential cross section $d\sigma
/d\Omega_\pi$, the right-hand-side (rhs) of Eq.\ (\ref{dcs4}) has to be integrated
over the value of the pion momentum $q=|\vec q|$ and the solid angle
$\Omega _{\vec P}$
\beqn
\label{dcs}
\frac {d\sigma}{d\Omega_\pi}=
\int_{q^{min}}^{q^{max}}\!\!\!\!q^2dq
\int d\Omega_{\vec  P}\frac {d\sigma }{d\vec q d\Omega_{\vec  P}}=
\frac 16~S,
\eeqn
where $S$ is defined as
\beq
S=\int_{q^{min}}^{q^{max}}\!\!\!\!fq^2dq
\int d\Omega_{\vec  P}
\sum _{m_2m_1\lambda m_d}
| \langle m_2m_1| T| \lambda m_d \rangle |^2,
\eeq
with
\beq
f=\frac 1{(2\pi)^5}\frac {m^2\varepsilon_d |{\vec P}|}
      {8k\cdot p_d~\varepsilon_\pi \varepsilon_P}.
\eeq
An extra factor of 1/2 must be included in the rhs of Eq.\
(\ref{dcs}) in case of charged pion photoproduction.
The maximum value $q^{max}$ can be found from Eq.\ (\ref{P_cm}) at
$W_{NN}=2m$. In the c.m. frame it is given by
\beqn
\label{q_cm}
q^{max}&=&\frac 1{2W_{\gamma d}}
\sqrt {[W^2_{\gamma d}-(2m+\mu)^2]
       [W^2_{\gamma d}-(2m-\mu)^2] },
\nn
q^{min}&=&0,
\eeqn where $\mu$ is the pion mass.
In the lab frame one has \beqn
\label{q_lab}
q^{max}=q^{max}(\Theta_\pi)&=&\frac 1b \left[ aE_\gamma
       z+(E_\gamma+M)\sqrt{a^2-b\mu^2}~\right] ,
\nn q^{min}=q^{min}(\Theta_\pi)&=&{\rm min}(0,\frac 1b \left[
aE_\gamma
       z-(E_\gamma+M)\sqrt{a^2-b\mu^2}~\right]) ,
\eeqn
 where $a=(W_{\gamma d}^2-4m^2+\mu^2)/2$ and
       $b=(E_\gamma+M)^2-E_\gamma^2z^2$ with $z=\cos{\Theta_\pi}$.
Note that the inequality $q^{min}\neq 0$ can take place only  for
$\Theta_\pi \le 90^\circ$, and at threshold energies
\beq E_\gamma
< E_\gamma^{max}=\frac { 4m^2-(M-\mu)^2}{2(M-\mu)}.
\eeq
The energy $E_\gamma^{max}$ is equal to 142.6 MeV
(if one takes $m=(m_p+m_n)/2=938.9$ MeV),
 149.0 MeV, and 146.2 MeV for $\pi^0$, $\pi^+$, and
$\pi^-$ channels, respectively. Therefore, in the
considered energy region, $q^{min}$ is equal to zero
also in the lab frame.

Apart from the differential cross section, single polarization
observables will be considered in the article, namely, the photon
beam asymmetry $\Sigma$ and target asymmetries $T_{IM}$.
Below, we give their definitions through the reaction amplitude
(see also Ref.~\cite{ArFix05}).
The photon asymmetry is
\begin{eqnarray}
\sum  =\frac {(d\sigma/d\Omega_\pi)^{\parallel}-
              (d\sigma/d\Omega_\pi)^{\perp}}
             {(d\sigma/d\Omega_\pi)^{\parallel}+
              (d\sigma/d\Omega_\pi)^{\perp}} =-
\frac {1}{S}~2
\int_{q^{min}}^{q^{max}}\!\!\!\!fq^2dq
\int d\Omega_{\vec  P}
\mbox{Re}\!\!\!\!\sum _{m_2m_1m_d}
\langle m_2m_1| T |+1m_d\rangle \langle m_2m_1| T | -1m_d\rangle ^*,
\label{asymm}
\end{eqnarray}
where $(d\sigma/d\Omega_\pi)^{\parallel (\perp)}$ is the inclusive cross
section for the photons polarized parallel (perpendicular)
to the $xz$-plane.
Note that the minus sign in the rhs of Eq.~(\ref{asymm})
is absent in the corresponding formulas from
Refs.~\cite{darwish04,darwish05_1,darwish05_2,darwish05_3,darwish05_4}.
The deuteron vector asymmetry $T_{11}$  and tensor asymmetries $T_{2M}$
are as follows
\footnote
{
The opposite sign
for $T_{11}$ is used   in
Refs.~\cite{darwish04,darwish05_4}.
}
\begin{eqnarray}
T_{11}&=&\frac {1}{S}~\sqrt {6}
\int_{q^{min}}^{q^{max}}\!\!\!\!fq^2dq
\int d\Omega_{\vec  P}~
\mbox{Im}\sum_{m_2 m_1\lambda} \left(
\langle m_2m_1| T | \lambda +1\rangle -
\langle m_2m_1| T | \lambda -1\rangle \right)
\langle m_2m_1| T | \lambda 0 \rangle ^*,
\nn
T_{20}&=&\frac {1}{S}~\frac {1}{\sqrt {2}}
\int_{q^{min}}^{q^{max}}\!\!\!\!fq^2dq
\int d\Omega_{\vec  P}~
\sum_{m_2m_1\lambda}
\left(| \langle m_2m_1| T | \lambda -1\rangle |^2+
| \langle m_2m_1| T | \lambda +1\rangle |^2-
2| \langle m_2m_1| T | \lambda 0\rangle |^2 \right),
\nn
T_{21}&=& \frac {1}{S}~\sqrt {6}
\int_{q^{min}}^{q^{max}}\!\!\!\!fq^2dq
\int d\Omega_{\vec  P}~
\mbox{Re}\sum_{m_2 m_1\lambda} \left(
\langle m_2m_1| T | \lambda -1\rangle -
\langle m_2m_1| T | \lambda +1\rangle \right)
\langle m_2m_1| T | \lambda 0 \rangle ^*,
\nn
T_{22}&=&\frac {1}{S}~2\sqrt {3}
\int_{q^{min}}^{q^{max}}\!\!\!\!fq^2dq
\int d\Omega_{\vec  P}~
\mbox{Re}\sum _{m_2m_1\lambda}
\langle m_2m_1| T | \lambda -1\rangle
\langle m_2m_1| T | \lambda +1\rangle ^*.
\label{t_IM}
\end{eqnarray}

\section{The theoretical model for inclusive pion photoproduction on
the deuteron}
\label{theory}

The diagrammatic approach is exploited to calculate
the amplitude $\langle m_2m_1| T| \lambda m_d \rangle $
in Eq.\ (\ref{dcs4}).
In comparison with Refs.~\cite{lps96,lsw00}, we reduce
the set of diagrams under consideration.
For example, in Ref.\ \cite{lsw00} where the
threshold region was considered, a two loop
diagram which includes simultaneously $np$ and $\pi N$ interactions
had to be taken into account.
Such a diagram is of importance at threshold energies because it
involves a block with charged pion photoproduction from the nucleon.
With  increasing photon energy this diagram becomes less important
as it was shown in Ref.\ \cite{lsw00}.
Above 200 MeV it can safely be disregarded.  It is known (see Refs.\
\cite{laget78,laget81}) that there are kinematic regions where a
one loop diagram with $\pi N$ rescattering noticeable contributes to
the amplitude. But this rather concerns the exclusive process $\gamma
d\to \pi NN$. We have checked that $\pi N$ rescattering changes the
final results  in the first resonance region by only a few percentages.

\begin{figure*}[hbt]
\includegraphics[width=0.6\textwidth]{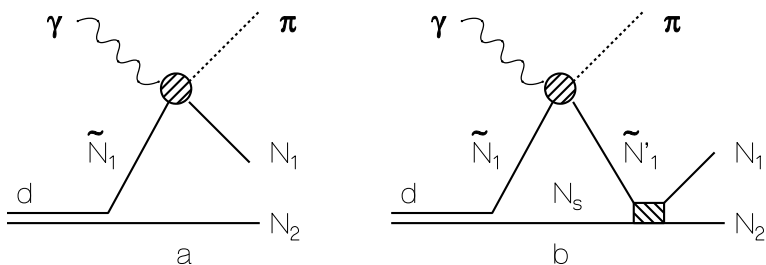}
\caption
{ Diagrams considered in this work. Two other diagrams
with the permutation $1\leftrightarrow 2$ are assumed. }
\label{fig1}
\end{figure*}

As a result, we retain in our calculations the two diagrams shown in
Fig.\ \ref{fig1}. The pole diagram \ref{fig1}(a) must be
taken into account
because at the integrations in Eq.\ (\ref{dcs}), one goes through the
kinematic regions where the relative momentum
$(\vec {\tilde p}_1-{\vec p}_2)/2$ is small
and, therefore, the deuteron wave function (DWF) has its maximum.
These are the so-called quasifree regions.
The exclusive cross section has sharp peaks in these regions.
Here the inclusive cross section from the pole diagrams is mainly saturated.
It is worth mentioning that in the peak regions the active
nucleon $\tilde N_1$ is almost on its mass shell.
In the center of the peaks the
difference between on-shell and off-shell energies of this nucleon is
equal to the deuteron binding energy $\Delta=2.2$ MeV. Therefore, the
use of the on-shell parametrization for a pion photoproduction
operator is justified when considering the diagram in Fig.~\ref{fig1}(a).
Nevertheless, as shown below, the off-shell dependence of
calculated observables does exist even when one considers
contributions from the pole diagram in Fig.~\ref{fig1}(a).

Another important mechanism is displayed in Fig.~\ref{fig1}(b).
When at the mentioned integration
the relative momentum of the outgoing nucleons,
$\vec p_{out}=(\vec p_2 - \vec p_1)/2$, decreases there are peaks in the
exclusive cross sections because of strong final state $NN$ interaction in the
$s$ waves (see, e.g., Refs.\ \cite{laget78,laget81,lps96}).
The peaks reveal themselves in
a big contribution of diagram \ref{fig1}(b) to
the inclusive cross section.  The effect of this diagram
is expected to be most
pronounced at small pion angles because in this case
the low-momentum regime simultaneously for both
DWF and $NN$ scattering amplitude is kinematically permitted.
The possibility of using the on-shell parameterization for
the pion photoproduction amplitude at evaluation of the diagram
\ref{fig1}(b) is less evident and it is discussed below.

Let us now write out the matrix elements corresponding
to the diagrams in Fig.\ \ref{fig1} (see also
Refs.\ \cite{laget78,laget81,lsw00,darwish03,FixAr05}).
One has for the pole diagram in Fig.~\ref{fig1}(a)
\beqn
\label{pole}
\langle m_2m_1| T^a(p_2,p_1,q;k)
| \lambda m_d\rangle =\sum _{\tilde m_1}
\Psi ^{m_d}_{m_2\tilde m_1}\left({\vec p}_2-\frac {\vec p_d}2
                               \right)
\langle m_1|
T_{\gamma {\tilde N_1}\to \pi N_1}(p_1,q;\tilde p_1,k)
| \lambda \tilde m_1 \rangle,
\eeqn
where $\Psi ^{m_d}_{m_2\tilde m_1}({\vec
p}_2-\vec p_d/2)$ is DWF and
$
\langle m_1|
T_{\gamma {\tilde N_1}\to \pi N_1}(p_1,q;\tilde p_1,k)
| \lambda \tilde m_1 \rangle$
is the amplitude of pion photoproduction on the nucleon.
There is one more pole diagram identical to that in
Fig.\ \ref{fig1}(a) but with the replacement
$1\leftrightarrow 2$.  In case of $\pi^0$ production the
corresponding matrix element should be added to Eq.\ (\ref{pole}).
For the charged channels a subtraction of  two matrix elements should
be done.

The deuteron wave function reads
\beqn
\label{dwf}
\Psi ^{m_d}_{m_2\tilde m_1}({\vec p})=(2\pi)^{3/2}\left[
\frac 1{\sqrt{4\pi}}
C^{1m_d}_{\frac 12 m_2 \frac 12 \tilde m_1}u(p)-
C^{1m_S}_{\frac 12 m_2 \frac 12 \tilde m_1}
C^{1m_d}_{2 m_L 1 m_S}Y^{m_L}_2( \hat {\vec p}) w(p)\right],
\eeqn
where $m_S=m_2+\tilde m_1$, $m_L=m_d-m_S$;
$Y^{m_L}_2( \hat {\vec p})$ are the spherical harmonics and
$C^{JM}_{J_2M_2J_1M_1}$ are the Clebsch-Gordan coefficients.  The
present calculation  is done using the CD-Bonn potential from
Ref.~\cite{mach01} where analytical parametrizations of the $s$ and $d$
amplitudes of DWF ($u(p)$ and $w(p)$, respectively) is given.
We note that our results are practically
independent of the choice of a model for $NN$-interaction.
Calculations with three OBEPR versions of the Bonn potential
\cite {mach87,mach89} or with a separable representation
\cite {PEST1} of the Paris potential give
almost the same predictions for observables.

If it is not stated otherwise, all the results below
are obtained with the production operator parametrized via
the invariant amplitudes $A_i$. They are defined in Eqs.~(\ref{TAi})
and (\ref{TGamma}) and calculated with both the SAID and MAID analyses.
The calculation for the SAID analysis is performed in two steps.
First, the CGLN amplitudes $F_i$ in the c.m. frame [Eq.~(\ref{CGLN})] are
found making the use of electric and magnetic multipoles predicted
by the analysis. We do not give explicit expressions
for $F_i$ through the multipoles because they are very well known.
Second, one finds the amplitudes $A_i$ using the relation
(\ref{AFamplitude}) between $A_i$ and $F_i$.
The MAID group provides users directly with the amplitudes $F_i$ for the
MAID00 solution and with both $F_i$ and $A_i$ for the MAID03
solution.

Strictly speaking there are other possible options for
the invariant amplitudes.
In particular, in Ref.~\cite{FixAr05} another set of those,
$A_i'$, as defined in Eqs.~(\ref{TAi'})
and (\ref{Ai'}) was exploited. The production operators
given by the amplitudes $A_i$ and $A_i'$ are
equivalent in the case of on-shell nucleons
as is explained in some detail in Appendix~\ref{operator}.
This equivalence  is destroyed when the nucleons are off
their mass shells. Because in deuteron calculations one deals
with off-shell nucleons, we expect our results to be dependent on
the parameterization of the elementary operator.

The matrix element corresponding to diagram \ref{fig1}(b) is
\beqn
\langle m_2m_1| T^b(p_2,p_1,q;k)
| \lambda m_d\rangle
=-m\int \frac {d^3\vec p_s}{(2\pi)^3}
\sum _{m_s\tilde m_1'}
\frac
{\langle{\vec p}_{out},m_2m_1| T_{NN}| {\vec p}_{in},
        m_s\tilde m_1'\rangle\langle
        m_s\tilde m_1'| T^a(p_s,\tilde p_1',q;k)
| \lambda m_d\rangle}
{p^2_{in}-p^2_{out}-i0}.
\label{np-resc}
\eeqn
The amplitude $\langle m_s\tilde m_1'|
T^a(p_s,\tilde p_1',q;k)| \lambda m_d\rangle $
in Eq.\ (\ref{np-resc}) is
the same as that in Eq.\ (\ref{pole}) but with the replacements
$2\to {\rm s}$ and $p_1\to \tilde p_1'$.
The second pole diagram mentioned above with
$1\leftrightarrow 2$ must also be included in the integrand of
Eq.\ (\ref{np-resc}). The choice of the energy of the off-shell
nucleon ${\tilde N}_1$ is discussed at the end of this section.

The half-off-shell $NN$ scattering amplitude
$\langle{\vec p}_{out},m_2m_1| T_{NN} |
{\vec p}_{in}, m_s\tilde m_1'\rangle$ depends on
the relative off-shell momentum  of the $N_1N_2$ pair before
scattering,
${\vec p}_{in} =\vec p_s-(\vec p_1+\vec p_2)/2$,
and the relative on-shell momentum after scattering,
$\vec p_{out}=(\vec p_2 - \vec p_1)/2$, as
\beqn
\langle{\vec p}_{out},m_2m_1| T_{NN} |
{\vec p}_{in}, m_{\rm s}\tilde m_1'\rangle
&=&(2\pi)^3
\sqrt{\frac {\varepsilon_{out}}m}
\sqrt{\frac {\varepsilon_{in} }m}
\sum _{JSLL'm_J}
C^{Sm_S}_{\frac 12 m_s \frac 12 \tilde m_1'}
C^{Sm'_S}_{\frac 12 m_2 \frac 12 m_1}
C^{Jm_J}_{Lm_LSm_S} C^{Jm_J}_{L'm_{L'}Sm'_S}    \nn  &&
\times ~i^{L-L'}
{Y^{m_L}_L}^*( \hat {\vec p}_{in})
Y^{m_{L'}}_{L'}( \hat {\vec p}_{out})R^{JS}_{L'L}(p_{out},p_{in}),
\label{np-ampl}
\eeqn
where $m_S=m_s+\tilde m_1'$, $m'_S=m_2+m_1$, $m_L=m_J-m_S$, and
$m_{L'}=m_J-m'_S$. The factors
$\sqrt{ \varepsilon_{out}/m}$ and $\sqrt{ \varepsilon_{in} /m}$
($\varepsilon_{out} =\sqrt {p^2_{out}+m^2}$ and
$\varepsilon_{in} =\sqrt {p^2_{in} +m^2}$ )
come from the so-called minimal relativity. The half-off-shell
partial amplitudes $R^{JS}_{L'L}(p_{out},p_{in})$ were obtained by
solving the Lippmann-Schwinger equation for the
CD-Bonn potential. The procedure for obtaining these amplitudes
is quite direct for  $np$ and $nn$ interactions.  It should be,
however, modified in the case of $pp$ interaction.
First, the Coulomb interaction has to be added to pure
nuclear interaction.
A method to handle Coulomb interaction  in momentum space was
proposed by  Vincent and Phatak \cite{vincent74}. We do not
discuss it here because it is described in full detail in that paper
(see also Refs.\ \cite{haidenb89,mach01,holz89}). We
mention only that the method was applied to the $^1S_0$ partial wave.
All other waves with $J=0$ and 1 are taken for the switched off
Coulomb potential. It makes no sense to include the Coulomb
modifications for the waves other than $^1S_0$ because even the
contribution of this latter to the observables was found to be
small. As a next step, we used
a prescription from Ref.\ \cite{kolybasov75} consisting
of the following parameterization of the half-off-shell
$^1S_0$ partial amplitude for $pp$ scattering
\beqn
R_{off}^{^1S_0}(p_{out},p_{in}) =
\frac {p_{out}^2+\beta^2} {p_{in}^2 +\beta^2}~
R_{on}^{^1S_0}(p_{out},p_{out}),
\eeqn
with $\beta=1.2$ fm$^{-1}$. The on-shell amplitude
$R_{on}^{^1S_0}(p_{out},p_{out})$ is obtained with the
use of the  Vincent and Phatak method with switched on
Coulomb interaction.

All partial waves with the total angular momentum $J\le 3$ were
retained in Eq.~(\ref {np-ampl}).  In fact, however, only
one wave, $^3S_1$, in the case $\pi^0$ photoproduction
is of importance. All other waves contribute give a few percentages
to observables. Further details of the
computations of Eq.~(\ref{np-resc}) can be found in Ref.\
\cite{lps96}.

\begin{figure*}[hbt]
\centerline{
\includegraphics[width=0.8\textwidth]{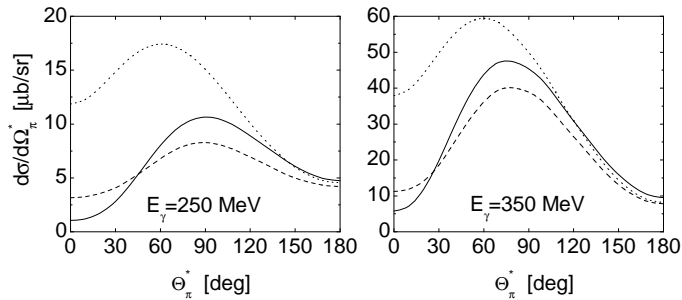}
}
\caption{
Differential cross section for the reaction $d(\gamma,\pi^0)np$ in
the c.m. frame at 250 and 350 MeV.
The dotted curves are the results without FSI.
Only the ``on-shell part" of the contribution of the diagram
in Fig.~\ref{fig1}(b) is retained in the dashed curves. Addition of
the ``off-shell part" gives the solid curves.
}
\label{fig2}
\end{figure*}

Again the question emerges whether  the on-shell parametrization for
the pion photoproduction amplitude is applicable at calculations of
the matrix element (\ref{np-resc}). To discuss this point, we
point out that at the evaluation of the integral in
Eq.\ (\ref{np-resc}) it is assumed, in accordance with a
finding from Refs.\ \cite{laget81,laget78}, the spectator nucleon
$N_s$ to be on its mass shell.  This means that the nucleon
$\tilde N_1$ momentum is $-\vec p_s$ and its energy is
equal to $2m-\Delta-\sqrt{p_s^2+m^2}$
(for simplicity we consider the lab frame).
The integral in Eq.\ (\ref{np-resc}) is saturated at momenta
$p_s \sim \sqrt{m\Delta}$ when DWF has its maximum.
In other words, the energy of the nucleon $\tilde N_1$ is
effectively off its on-shell value by only few multiplicities
of binding energies $\Delta$.
Furthermore, using the symbolic equality
\beqn
\label{equal}
\frac 1{p^2_{in}-p^2_{out}-i0}=
\frac {i\pi} {2p_{out}}\delta(p_{in}-p_{out})+
P~\frac 1{p^2_{in}-p^2_{out}},
\eeqn
one can split the matrix element (\ref{np-resc}) in its on-shell
part and its off-shell part corresponding to the first and second
terms in the rhs  of Eq.\ (\ref{equal}), respectively.  With only
the former included we calculated the cross sections and found it to
give the main contribution (see Fig.~\ref{fig2}).  Taking into
account that the nucleon $N_s$ is on its mass shell, one
concludes that this part corresponds to the case when the nucleon
$\tilde N_1'$ is also on its mass shell.  Therefore, the
contribution of the amplitude Eq.\ (\ref{np-resc}) to
the inclusive cross section comes mainly from the kinematic domains
in the integrand of Eq.\ (\ref{np-resc}) where the nucleons $\tilde N_1$
and $\tilde N_1'$ are close to their mass shells. The same conclusion
holds true also for other observables. Therefore, the
on-shell parameterization for an elementary pion photoproduction
operator is applicable in this integrand. Nevertheless, some
dependence of the FSI amplitude [Eq.~(\ref{np-resc})] on off-shell effects
is expected.

As in Refs.~\cite{lps96,lsw00},
all summations over polarizations of the particles in Eqs.\
(\ref{pole}) and (\ref{np-resc}) as well as the three-dimensional
integration in Eq.\ (\ref{np-resc}) have been carried out
numerically. The number of chosen nodes at this integration and that
in Eq.\ (\ref{dcs}) was taken to be sufficient for prediction of
observables with the numerical accuracy better than 2\%.

\section{Results and Discussion}
\label{results}

\subsection{Differential and total cross sections}

\begin{figure*}[tbh]
\includegraphics[width=0.6\textwidth]{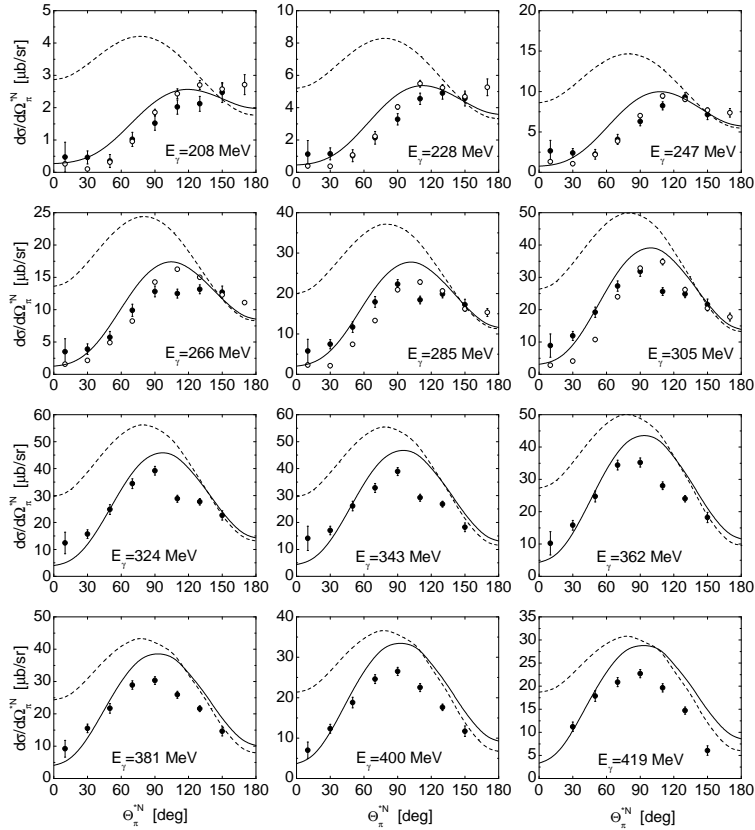}
\caption
{
Differential cross section for $\pi^0$ production
in the photon-nucleon c.m. frame
obtained with the FA04K solution of the SAID analysis
and with the parametrization of the operator
through the invariant amplitudes $A_i$.
The dotted (solid) curves are our predictions without (with) FSI.
Data are from Ref.\ {\protect \cite{krusche99}} ($\bullet$) and
Ref.\ {\protect \cite{siodl01}} ($\circ$).
}
\label{fig3}
\end{figure*}

We begin our discussion with the results for the neutral channel.
In Fig.\ \ref{fig3}, the predicted differential cross
sections of $\pi^0$ production are shown in the energy region
between 208 and 419 MeV together with experimental results
from  Refs.~\cite{krusche99,siodl01}
\footnote
{
In Refs.\ \cite{krusche99,siodl01} the differential cross sections
are given in the so-called ``photon-nucleon c.m. frame''. Relations
needed to transform the cross sections and angles from  the $\gamma
d$ c.m. frame to the frame mentioned are presented in
Refs.~\cite{LSW00,darwish_diss}.
}.
The displayed cross sections are obtained
with a pion photoproduction operator
parameterized through the amplitudes $A_i$ and the SAID FA04K solution.
One can see one more confirmation of a prediction from
Refs.\ \cite{laget81,lps96} that the effect of $np$ final state
interaction should lead to a reduction of the cross section
and this reduction is the stronger the smaller the pion angles are.
This effect is mainly attributed to the strong $np$
interaction in the $^3S_1$ wave.

Without FSI the model completely  fails to reproduce the data.
After including FSI, the curves move to the data points
although a reasonable description of them still remains to be achieved.
At $60^\circ \leq  \Theta^{*N}_\pi \leq 120^\circ$, the predicted
cross sections overestimate the data by about $10-20\%$.

Possible explanations of the disagreement between the data
and the present model can be looked for in the elementary
photoproduction operator. As is explained above, the on-shell
parameterization for the latter can be used when studying
the inclusive channels. Nevertheless, different representations
of the operator, which are equivalent in the on-shell case,
turn out to be not quite equivalent when one or two nucleons
are off their mass shells.
Because the sophisticated phenomenological analyses
like MAID or SAID provide the elementary amplitude
for the on-shell nucleons only, it is hardly possible to give
precise quantitative account for the
off-shell effects using these amplitudes.
One can, however, estimate the possible size of
the off-shell effects by performing calculations with different
representations of the operator. As an example, we make use of
two forms of the operator given in Appendix \ref{operator},
corresponding to the amplitudes $A_i$ (\ref{TAi}) and $A_i'$ (\ref{TAi'}).
In addition, the MAID03 solution was used to parametrize
this operator.

\begin{figure*}[hbt]
\includegraphics[width=0.8\textwidth]{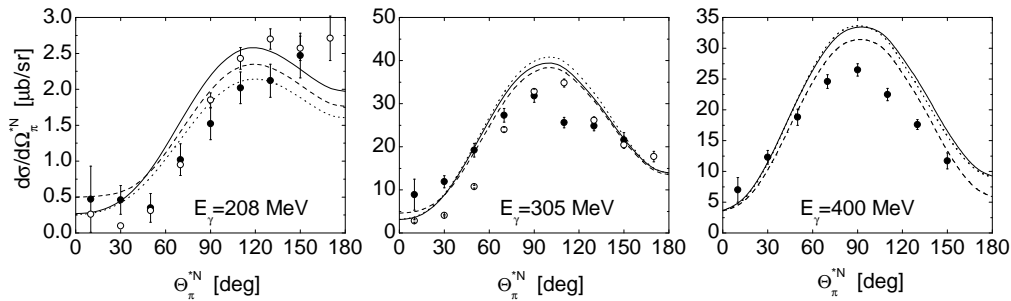}
\caption
{
Differential cross section for the $\pi^0$ channel
in the photon-nucleon c.m. frame
with different parametrizations of the
production operator: dashed and solid curves
are obtained  with the amplitudes
$A_i'$ and $A_i$, respectively, for the SAID FA04K solution.
Dotted curves are obtained with the amplitudes $A_i$ and
the MAID03 solution.
Data described in the legend to Fig.~\ref{fig3}.
}
\label{fig4}
\end{figure*}

The cross sections, shown in Fig.~\ref{fig4},
exhibit quite noticeable dependencies to the different
parametrizations of the operator. At 208 MeV,
the cross section is sensitive both to the choice of
the analysis and the on-shell form of the operator. With
increasing photon energy, the sensitivity to the
analysis diminishes but the sensitivity to the
form of the operator remains to be noticeable. The regions
overlapped by the curves can be considered as characterizing
the size of possible uncertainties introduced by the pion
photoproduction operator. They should be kept in mind at
attempts to extract the cross sections on the neutron from
deuteron data.

\begin{figure*}[hbt]
\includegraphics[width=0.8\textwidth]{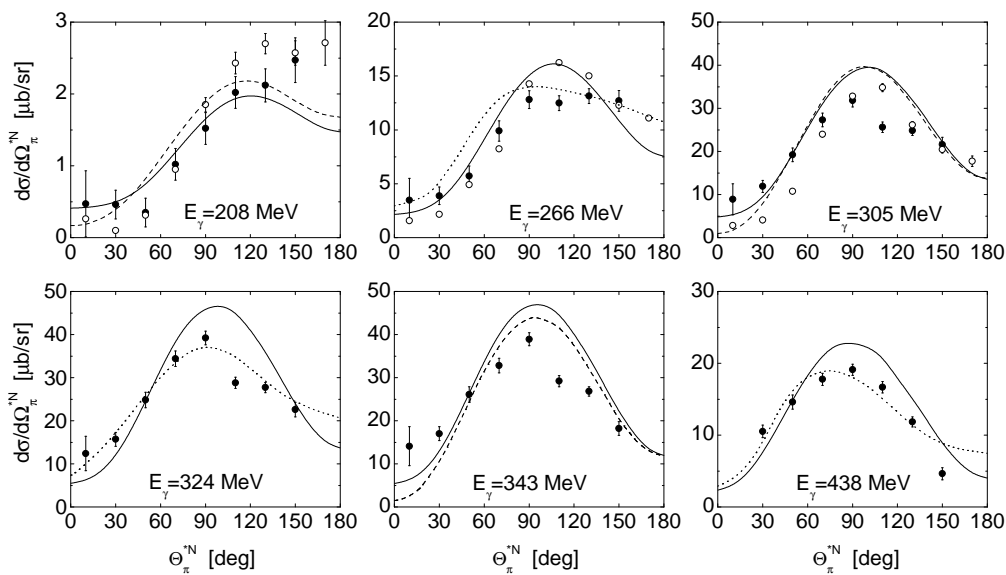}
\caption
{
Differential cross section for the reaction
$d(\gamma,\pi^0)np$ in the photon-nucleon c.m. frame
obtained in three models: the present calculation
with the amplitudes $A_i'$ and the MAID03 solution (solid),
Ref.~\cite{darwish03} (dotted), and Ref.~\cite{FixAr05} (dashed).
Data described in the legend to Fig.~\ref{fig3}.
}
\label{fig5}
\end{figure*}

\begin{figure*}[hbt]
\includegraphics[width=0.8\textwidth]{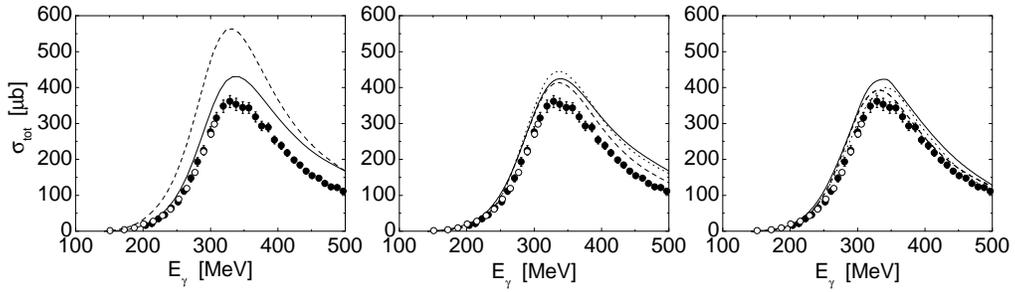}
\caption
{
Total cross section for the reaction
$d(\gamma,\pi^0)np$.
(Left panel)
Results with the FA04K solution of the SAID analysis
and with the parametrization of a pion production operator
through the invariant amplitudes $A_i$. The dotted (solid) curves
are the predictions without (with) FSI.
(Middle panel)
Dashed and solid curves
are obtained  with the amplitudes
$A_i'$ and $A_i$, respectively, for the SAID FA04K solution.
The dotted curve  is obtained with the amplitudes $A_i$ and
the MAID03 solution.
(Right panel)
The present calculation
with the amplitudes $A_i'$ and the MAID03 solution (solid),
Ref.~\cite{darwish03} (dotted), and Ref.~\cite{FixAr05} (dashed).
Data described in the legend to Fig.~\ref{fig3}.
}
\label{fig6}
\end{figure*}

A comparison of  our results with those from recent works
\cite{darwish03} and \cite{FixAr05} is presented in Fig.~\ref{fig5}.
Because in the latter article a photoproduction
operator was parametrized via the amplitudes $A_i'$ and
the MAID03 solution, we give the comparison with the same operator.
One can see a satisfactory agreement with the results from Ref.~\cite{FixAr05}.
Slight deviation might be attributed to the use of the different
parametrizations for the half-off-shell $NN$ scattering amplitude.
At the same time, our differential cross section exhibits quite a
different behavior compared to that from Ref.~\cite{darwish03}.
Reasons responsible for this incongruity of the results can be
in the use of both different half-off-shell $NN$ scattering amplitudes
and elementary photoproduction operators. We suppose the latter reason
to be more probable.

\begin{figure*}[hbt]
\includegraphics[width=0.6\textwidth]{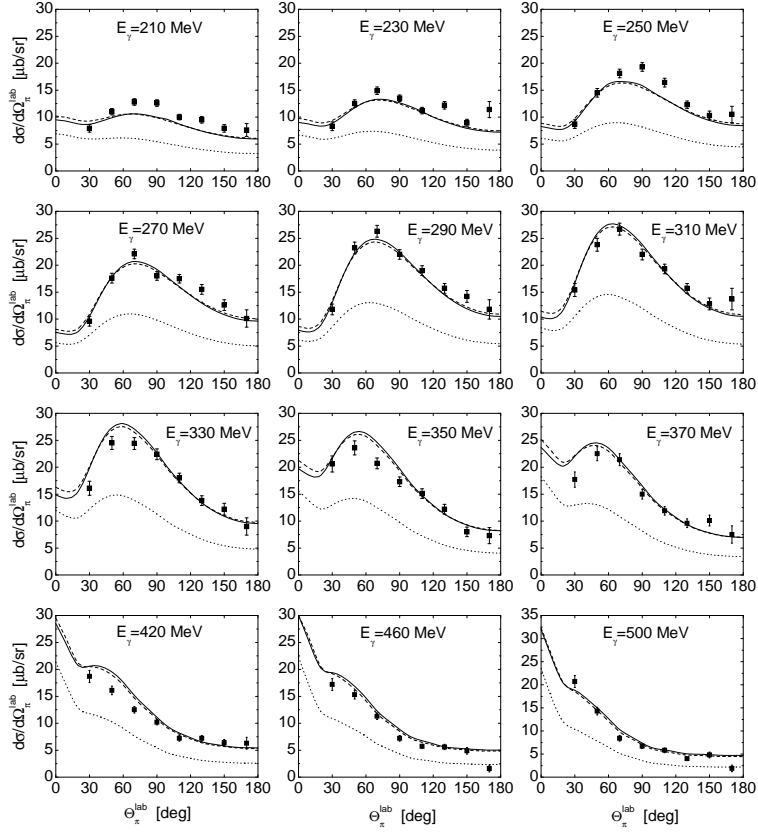}
\caption{ Differential cross section for the reaction
$d(\gamma,\pi^-)pp$ in the lab frame
obtained  with the amplitudes $A_i$ and the SAID FA04K solution.
The dotted curves are
contributions from one of the pole diagrams in Fig.\ \ref{fig1}.
Successive addition of the second pole diagram and FSI leads to
dashed and solid curves, respectively.
Data described in the legend toRef.~\cite{benz73}.
 }
\label{fig7}
\end{figure*}

After integrating Eq.\ (\ref{dcs}) over the solid pion angle one
obtains the total cross section for a given channel. In Fig.~\ref{fig6},
the total cross section for $\pi^0$ photoproduction is
shown. It is clear that everything told above on the differential
cross section also holds true for the total cross section. In particular,
one can see that the model without FSI clearly overestimates the data.
Inclusion of FSI strongly reduces the cross section although
it still noticeable overestimate the data from Refs.~\cite{krusche99}
and \cite{siodl01} for all parametrizations of an elementary photoproduction
operator.  Our total cross sections
are about 5\% higher as compared to those from Ref.~\cite{FixAr05}
in the vicinity of the peak. Note also that in comparison to
the prediction from Ref.~\cite{darwish03}, our peak position
is shifted on about 10 MeV to smaller energies that seemingly is because
different elementary photoproduction operators used in two models.

\begin{figure*}[hbt]
\includegraphics[width=0.8\textwidth]{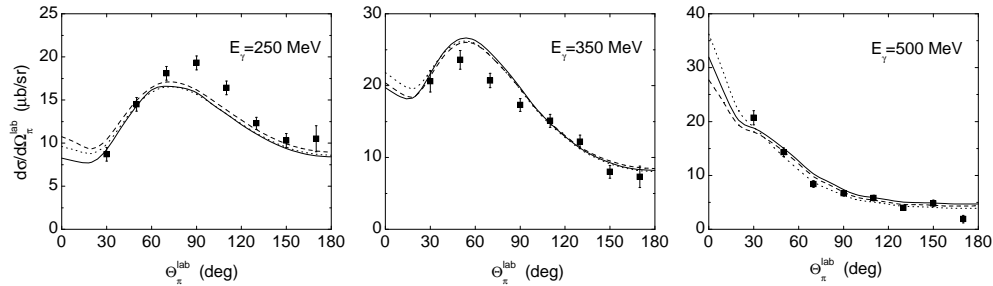}
\caption
{
Differential cross section for the reaction
$d(\gamma,\pi^-)pp$ in the lab frame
with different parametrizations of a
production operator. Notations of the curves as in Fig.~\ref{fig4}.
Data are from Ref.~\cite{benz73}.
}
\label{fig8}
\end{figure*}

In discussion of charged pion photoproduction we
restrict ourselves to the $d(\gamma,\pi^-)pp$ channel only
because results for the $d(\gamma,\pi^+)nn$ channel are very similar.
The dotted curves in Fig.~\ref{fig7} that
correspond to the contribution of one pole diagram, reproduce the
behavior of the angular dependence for the differential cross
section of the elementary reaction $\gamma n\to\pi^-p$. In
particular, at energies above 400 MeV, a sharp peak at forward
angles because of the pion exchange in the $t$ channel is clearly seen.
One can see that at $\Theta_\pi\geq 90^\circ$ the cross section from
two pole diagrams is practically equal to twice the cross
section from one diagram. The reason for this is that at
backward angles
the events where both nucleons have small momenta correspond to the high
momentum components of the deuteron wave function
and, therefore, both diagrams cannot
work  in the quasifree regime at the same time.
As a result, the interference term is very small for backward angles.
Of course, this conclusion
is valid for all channels. In full agreement with a finding from
Refs.~\cite{darwish03,FixAr05},
the effect from FSI has only a marginal impact on
the differential cross section.

As is seen in Fig.~\ref{fig8}, the sensitivity of the cross section
to the different parametrizations of the photoproduction operator is not
as strong as in the case of the $\pi^0$ channel. It remains to be
visible only at forward angles.

\begin{figure*}[hbt]
\includegraphics[width=0.8\textwidth]{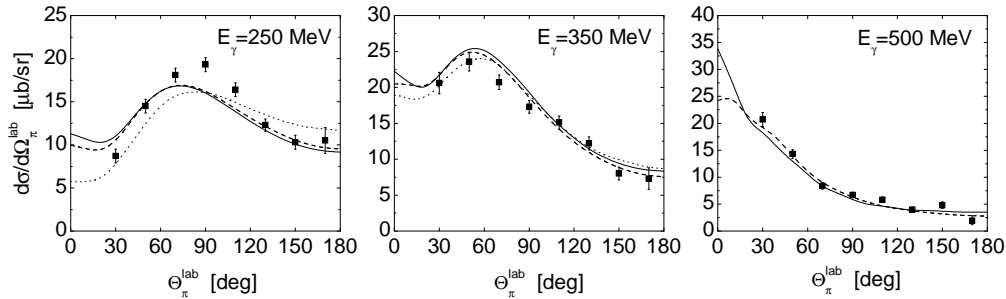}
\caption
{
Differential cross section for the reaction
$d(\gamma,\pi^-)pp$ in the lab frame
obtained in three models:
the present calculation
with the amplitudes $A_i'$ and the MAID03 solution (solid),
Ref.~\cite{darwish03} (dotted), and Ref.~\cite{FixAr05} (dashed).
Data are from Ref.~\cite{benz73}.
}
\label{fig9}
\end{figure*}

A comparison of  our results with those from recent works
\cite{darwish03,FixAr05} is presented in Fig.~\ref{fig9}.
One can see the good agreement with the results from Ref.~\cite{FixAr05}.
We, however, have expected better agreement because the FSI effect
is small for the charged channels and, as stated
in Sect.~\ref{theory}, the results are independent of a choice of DWF.
The deviation at forward angles is even somewhat bigger
because the Coulomb forces between the protons
have been disregarded in Ref.~\cite{FixAr05}. Their effect consists in
the decrease of the cross section on $5-10\%$ at zero angle
in the energy region from 250 to 500 MeV and
becomes to be negligible at $\Theta_\pi\agt 30^\circ$.
Therefore, reasons for the slight deviations between our
results and these from Ref.~\cite{FixAr05} remain to be investigated.
The disagreement with the results from Ref.~\cite{darwish03} is
also difficult to explain because, as it is stated in
Ref.~\cite{darwish03}, the operator used in that work provides the
differential cross sections of the elementary reaction on the
nucleon close to those given by the MAID analysis.

\begin{figure*}[hbt]
\includegraphics[width=0.8\textwidth]{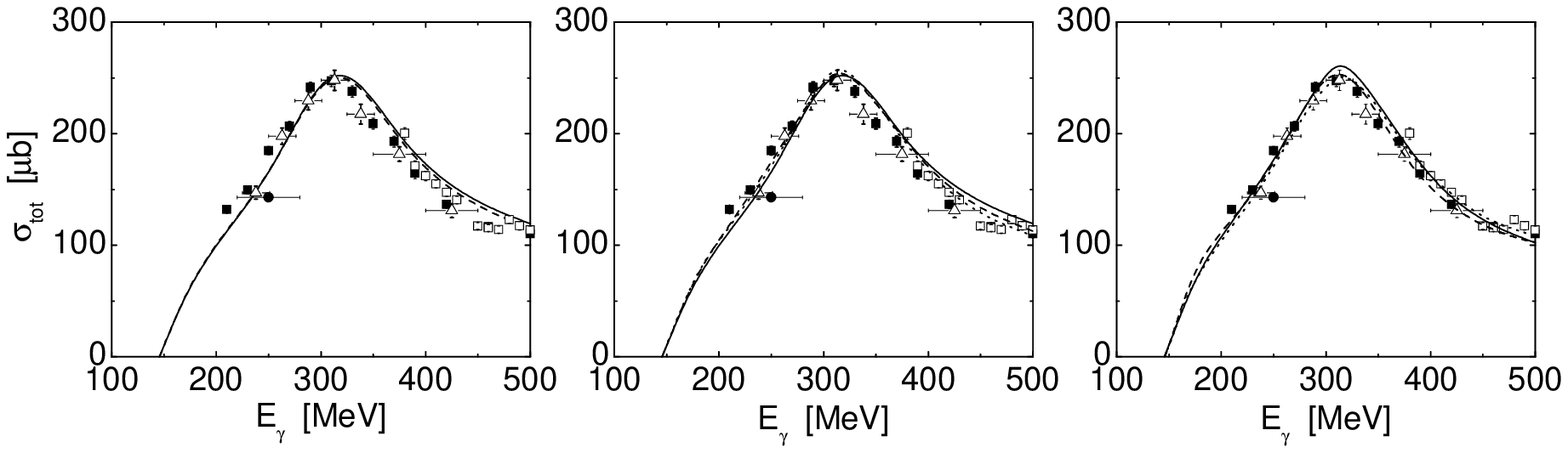}
\caption
{
Total cross section for the reaction
$d(\gamma,\pi^-)pp$. Notations of the curves as
in Fig.~\ref{fig6}.
Data are from Refs.~\cite{benz73} ($\blacksquare$),
\cite{chief75} ($\triangle$), \cite{asai90} ($\Box $), and
\cite{quraan98} ($\bullet$).
}
\label{fig10}
\end{figure*}

The total cross section for $\pi^-$ photoproduction is shown in
Fig.~\ref{fig10}. Here the FSI contribution is much smaller
than that for $\pi^0$ production and leads to a slight decrease of the
cross section above 350 MeV. We find satisfactory agreement with data
from Refs.~\cite{benz73,chief75,asai90}. At the same time a data point
from Ref.\ \cite{quraan98} at 250 MeV lies markedly below both
our predictions and data from Refs.~\cite{benz73,chief75}.
The sensitivity of the results to the choice of the photoproduction
operator is rather small.
A comparison to the results from Refs.~\cite{darwish03,FixAr05}
shows that three models give very similar results.

Having results for the total cross sections in all the channels
mentioned above one can try to make predictions for the total
photoabsorption cross section on the deuteron in the first resonance
region.  Of course, two more reactions contribute to it as well.
These are coherent $\pi^0$ photoproduction from the deuteron,
$\gamma d\to\pi^0 d$, and deuteron photodisintegration, $\gamma d\to
np$.  Predictions for the former are taken from a model built in
Ref.~\cite{kamalov97} which provides a good description of
data from Ref.~\cite{krusche99}. The total cross section for
the latter reaction is calculated making use of a phenomenological
fit \cite{ross89} to available experimental data on deuteron
photodisintegration up to 440 MeV.

\begin{figure*}[hbt]
\includegraphics[width=0.8\textwidth]{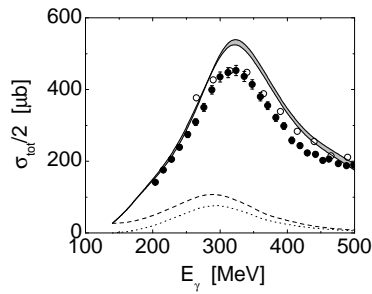}
\caption{
Total photoabsorption cross section per nucleon for the deuteron from
150 to 500 MeV.  Contribution of the reaction $\gamma d\to\pi^0 d$
is shown as dotted curve. Contribution from deuteron photodisintegration
is included in dashed curve. Filled area includes contributions
from  all channels $\gamma d\to \pi NN$ (see text).
Data are from Refs.\ \cite{armstr72} ($\circ$)
and \cite{maccorm96} ($\bullet$).
}
\label{fig11}
\end{figure*}

In Fig.\ \ref{fig11} we present our results for the total
photoabsorption cross section per nucleon for the deuteron. The filled
area includes the uncertainties discussed above, because os the
variations of the elementary photoproduction operator.
It is seen that the predictions even with allowance for
these uncertainties are noticeable above the data from Refs.\
\cite{armstr72,maccorm96} in the peak region.
In the center of the peak at about 320 MeV we find
our prediction of $(543\pm 7)~\mu b$ strongly overestimating the
experimental value of $(452\pm 5)~\mu b$.
We have no explanation for this disagreement. There are reasons
for the assumption that the total cross sections from
Refs.~\cite{armstr72,maccorm96} may be too low.
Such an assumption is supported by the study of
other electromagnetic processes. As an example, one can mention
that the sum of the electric and magnetic polarizabilities
of the neutron calculated with those cross sections turns out to
be notably underestimated (see Ref.~\cite{LL00} for a more detailed
discussion).

\subsection{Beam asymmetry for linearly polarized photons}

\begin{figure*}[hbt]
\includegraphics[width=0.6\textwidth]{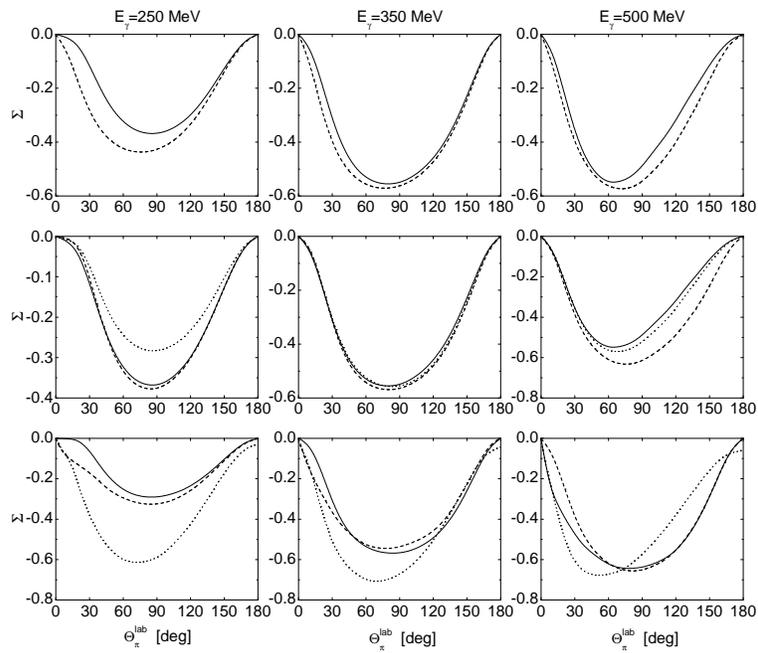}
\caption
{
Angular distribution of the asymmetry $\Sigma$ for the reaction
$d(\gamma,\pi^0)np$.
(Upper panel)
The dotted (solid) curves are our predictions without (with) FSI
obtained with the SAID FA04K solution and with the amplitudes $A_i$.
(Middle panel)
Dashed and solid curves
are obtained  with the amplitudes
$A_i'$  and $A_i$, respectively, for the SAID FA04K solution.
Dotted curves are obtained with the amplitudes $A_i$ and
the MAID03 solution.
(Lower panel)
The present calculation with the amplitudes $A_i'$
and the MAID03 solution (solid),
Refs.~\cite{darwish04,darwish05_1} (dotted), and
Ref.~\cite{FixAr05} (dashed).
}
\label{fig12}
\end{figure*}

The beam asymmetry $\Sigma$ for $\pi^0$ production at three
selected energies of 250, 350, and 500 MeV is displayed
in Fig.~\ref{fig12}. In IA it is negative at all energies.
FSI results in a decrease of the magnitude of $\Sigma$.
The influence from FSI is noticeable at the lowest
energy and forward angles. With increasing energy the effect
of FSI becomes smaller although not negligible even
at the highest energy.

As in case of the differential cross section, the beam
asymmetry depends strongly on the form
of the elementary production operator. As is seen
in Fig.~\ref{fig12}, at 250 MeV there exists
noticeable sensitivity of $\Sigma$ to the choice of
the analysis of photomeson amplitudes and to the choice of their
representations in terms of $A_i$ or $A_i'$. This sensitivity
mainly reflects the difference between the parametrizations
FA04K and MAID03  in case of $\pi^0$ production near 250 MeV.
In the $\Delta$ region
and at higher energies, the asymmetry is practically
independent of the analysis. But here the influence of the choice of
the amplitudes $A_i$ or $A_i'$ becomes visible.
Only at 350 MeV these amplitudes lead to very close results for
the asymmetry. From this finding one may conclude
that the $\Delta$ region is promising for a model
independent determination of $\Sigma$ for
the $\pi^0n$ channel from deuteron data. Outside this region, results
of such an extraction will be strongly model
dependent.

A comparison of our results with those from
Refs.~\cite{darwish04,darwish05_1} and \cite{FixAr05}
is also presented in Fig.~\ref{fig12}. Our predictions are
similar to those from the latter work although
there is a disagreement at forward angles for 250 MeV. As
to the predictions from Ref.~\cite{darwish04,darwish05_1},
we have a notable disagreement
with the results from those works in absolute size of $\Sigma$
at 250 and 350 MeV and in the form of the
angular distribution at 500 MeV.
One should mention a strange result of
Refs.~\cite{darwish04,darwish05_1} consisting
in the statement that $\Sigma$ does not vanish at $\Theta_\pi=0$
and $\pi$, as it has to be because of helicity
conservation~\cite{ArFix05,FixAr05}.

\begin{figure*}[hbt]
\includegraphics[width=0.8\textwidth]{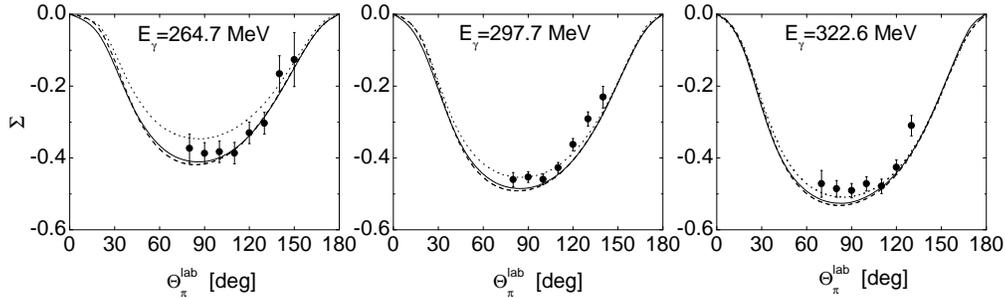}
\caption
{
Angular distribution of the asymmetry $\Sigma$ for the reaction
$d(\gamma,\pi^0)np$. Notation of the curves as in the middle
panel of  Fig.~\ref{fig12}.
Data (preliminary) are from Ref.~\cite{legs_np}.
}
\label{fig13}
\end{figure*}

Recently first preliminary data on the asymmetry $\Sigma$ in the
$\pi^0$ and $\pi^-$ channels at a few energies between
265 and 330 MeV have been reported by the LEGS collaboration
\cite{legs_np}. A comparison with the data for $\pi^0$ production
is presented in Fig.~\ref{fig13}. One can readily see a satisfactory
agreement with the experimental values for all parametrizations of
the photoproduction operator. Only near $90^\circ$ the asymmetry
is slightly overestimated in absolute size when using the operator built
with the SAID model.

\begin{figure*}[hbt]
\includegraphics[width=0.6\textwidth]{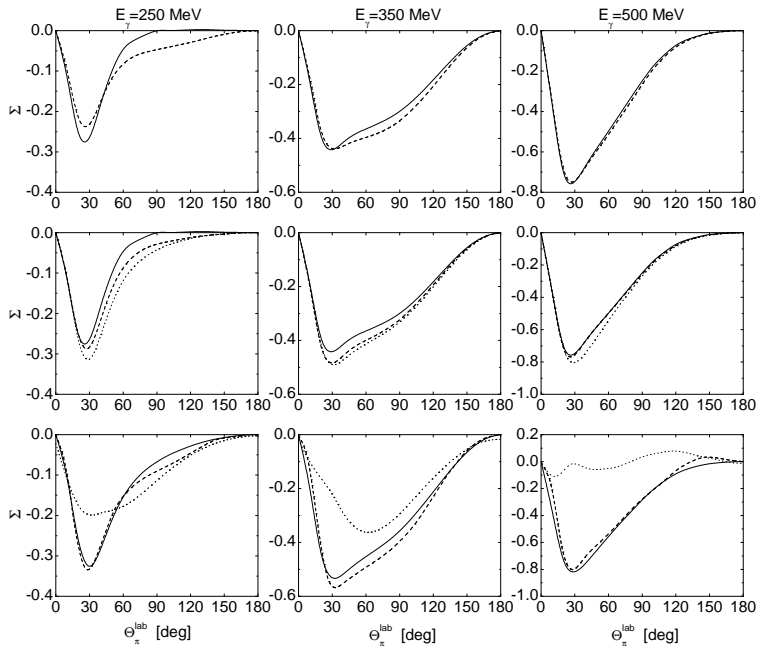}
\caption
{
Angular distribution of the asymmetry $\Sigma$ for the reaction
$d(\gamma,\pi^-)pp$. Notation of the curves as in Fig.~\ref{fig12}.
Results from Refs.~\cite{darwish05_3,darwish05_4} are shown
as dotted curves of the lower panel.
}
\label{fig14}
\end{figure*}

The beam asymmetry $\Sigma$ for $\pi^-$ production is
shown in Fig.~\ref{fig14}.  Remaining negative as in
the case of $\pi^0$ production, it shows a quite
different behavior. One can see a sharp peak near
$\Theta_\pi \simeq 30^\circ$.
The effect from FSI is much smaller and has a noticeable impact on
$\Sigma$ only at the lowest energy. At the highest energy
it is negligible.
The SAID and MAID analyses give quite different results
for $\Sigma$ at the lowest energy.
This difference is diminishing when the energy increases.
Notable influence on the predictions at 250 MeV has also
the form of the production operator.

In the same figure we compare our predictions to
those from Refs.~\cite{darwish05_3,darwish05_4,FixAr05}.
Note that polarization observables for the $\pi^-$ channel
were calculated also in Refs.~\cite{darwish04,darwish05_1}.
In some cases they are in notable disagreement with
those given in Refs.~\cite{darwish05_3,darwish05_4}.
Authors do not explain reasons for the deviation. In this
situation we preferred to make a comparison with
predictions from the more recent works.
Our results are in reasonable agreement with the predictions of
Ref.~\cite{FixAr05}, but in substantial disagreement with
the results of Refs.~\cite{darwish05_3,darwish05_4}
both for the form of the angular distribution and
for the absolute size of $\Sigma$
\footnote
{In Refs.~\cite{darwish05_3,darwish05_4} the angular
distribution of $\Sigma$  is
given at $E_\gamma=200$,
270, 330, 370, 420, and 500 MeV. The curves at 250 and 350 MeV
displayed in Fig.~\ref{fig14}
have been obtained by a quadratic interpolation. The same
procedure is used to obtain the target asymmetries $T_{IM}$
for $\pi^-$ production discussed below.
}.
The disagreement is drastic at the highest energy and
can hardly be caused by the use of a different elementary
production operator. It is likely, that there is an unnoticed
computational error in Refs.~\cite{darwish05_3,darwish05_4},
leading also to the odd results of nonzero asymmetry
at $\Theta_\pi=0$ and $\pi$.

\begin{figure*}[hbt]
\includegraphics[width=0.8\textwidth]{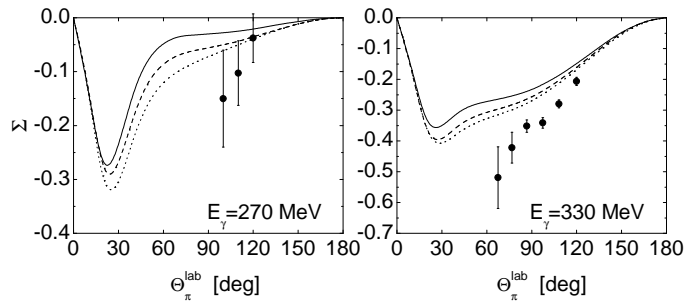}
\caption
{
Angular distribution of the asymmetry $\Sigma$ for the reaction
$d(\gamma,\pi^-)pp$. Notation of the curves as in the middle panel
of Fig.~\ref{fig14}.
Data (preliminary) are from Ref.~\cite{legs_np}.
}
\label{fig15}
\end{figure*}

In Fig.~\ref{fig15} we compare our results with the preliminary
data from the LEGS collaboration at 270 and 330 MeV. Although some model
dependence of the predictions exists, the calculated asymmetry
is too small in its absolute size at all the parametrizations
of the production operator.

We do not consider here the $\pi^+$
channel because all the conclusions just drawn for $\pi^-$
production remain to be valid for $\pi^+$ production as well
(see also Ref.~\cite{FixAr05}) and move to a discussion of
the target asymmetries.

\subsection{Target asymmetries for polarized deuterons}

\begin{figure}[hbt]
\includegraphics[width=0.6\textwidth]{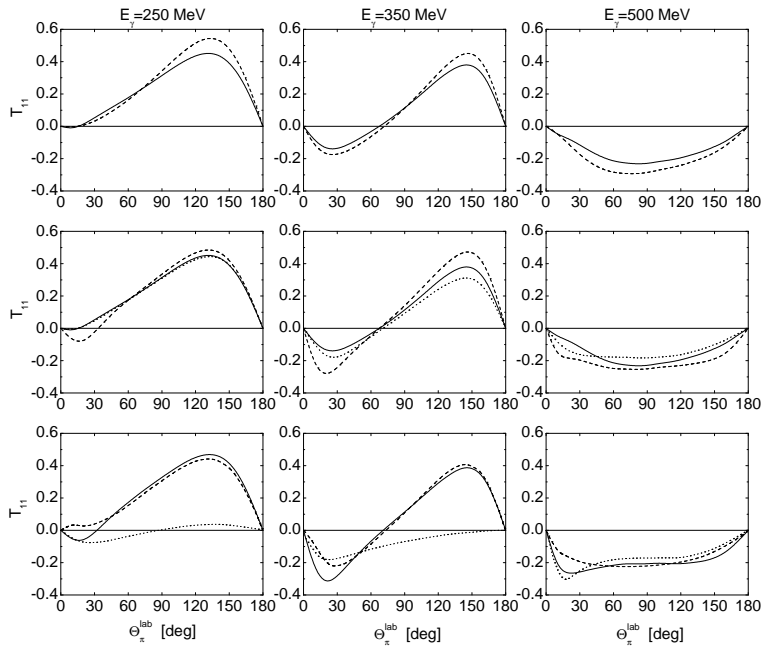}
\caption
{
Target asymmetry $T_{11}$ for the reaction $d(\gamma,\pi^0)np$.
Notation described in the legend to Fig.~\ref{fig12}.
}
\label{fig16}
\end{figure}

\begin{figure}[hbt]
\includegraphics[width=0.6\textwidth]{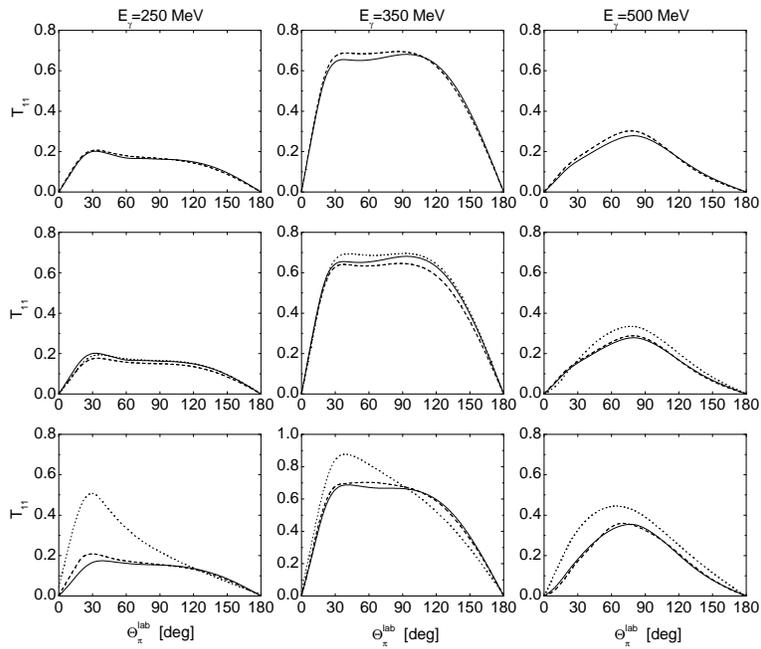}
\caption
{
Target asymmetry $T_{11}$ for the reaction $d(\gamma,\pi^-)pp$.
Notation of the curves as in Fig.~\ref{fig14}, only in the lower
panel, results from Ref.~\cite{darwish05_4} are
shown in dotted curves.
}
\label{fig17}
\end{figure}

\begin{figure}[hbt]
\includegraphics[width=0.6\textwidth]{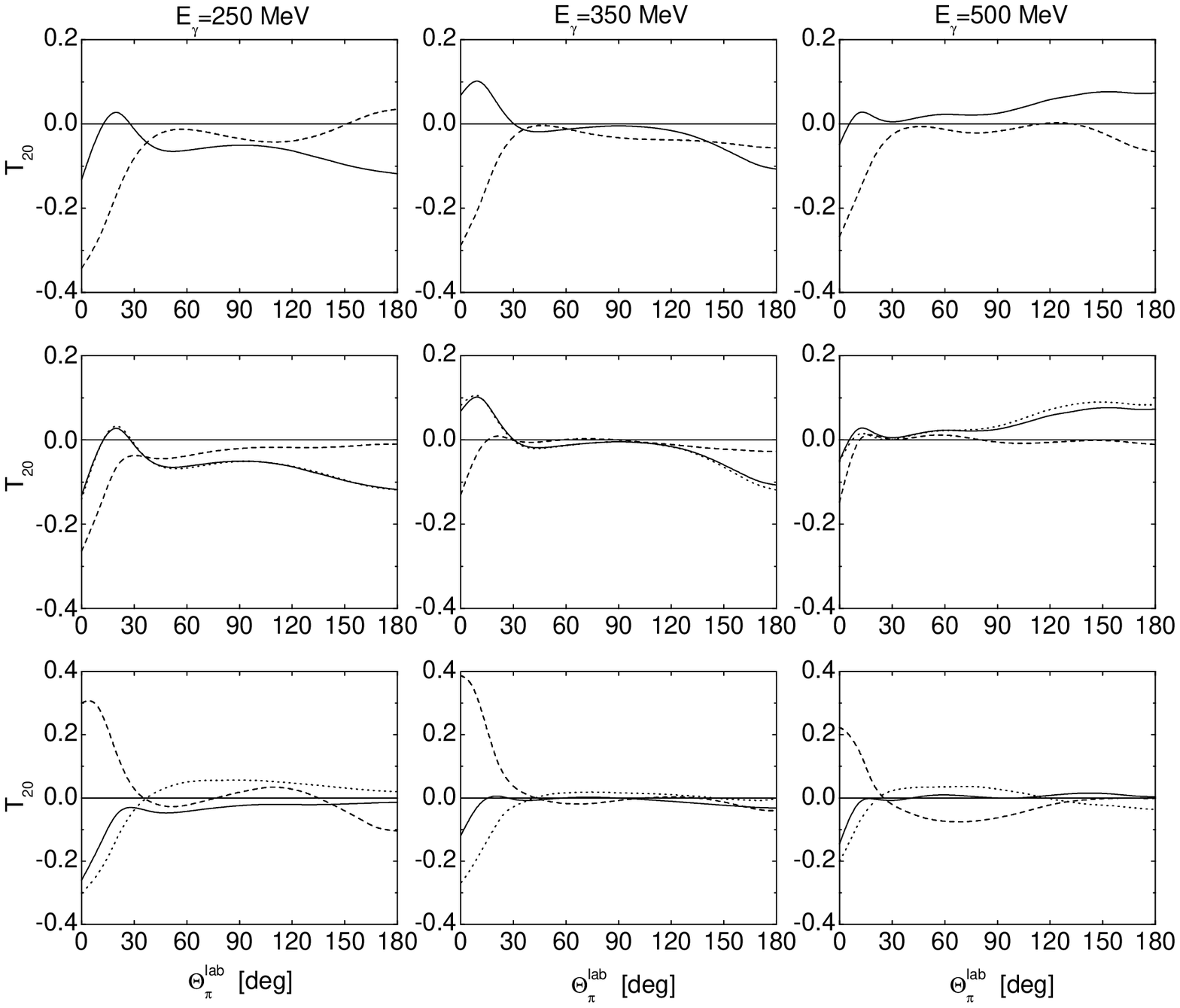}
\caption
{
Target asymmetry $T_{20}$ for the reaction $d(\gamma,\pi^0)np$.
Notation described in the legend to Fig.~\ref{fig12}.
}
\label{fig18}
\end{figure}

\begin{figure}[hbt]
\includegraphics[width=0.6\textwidth]{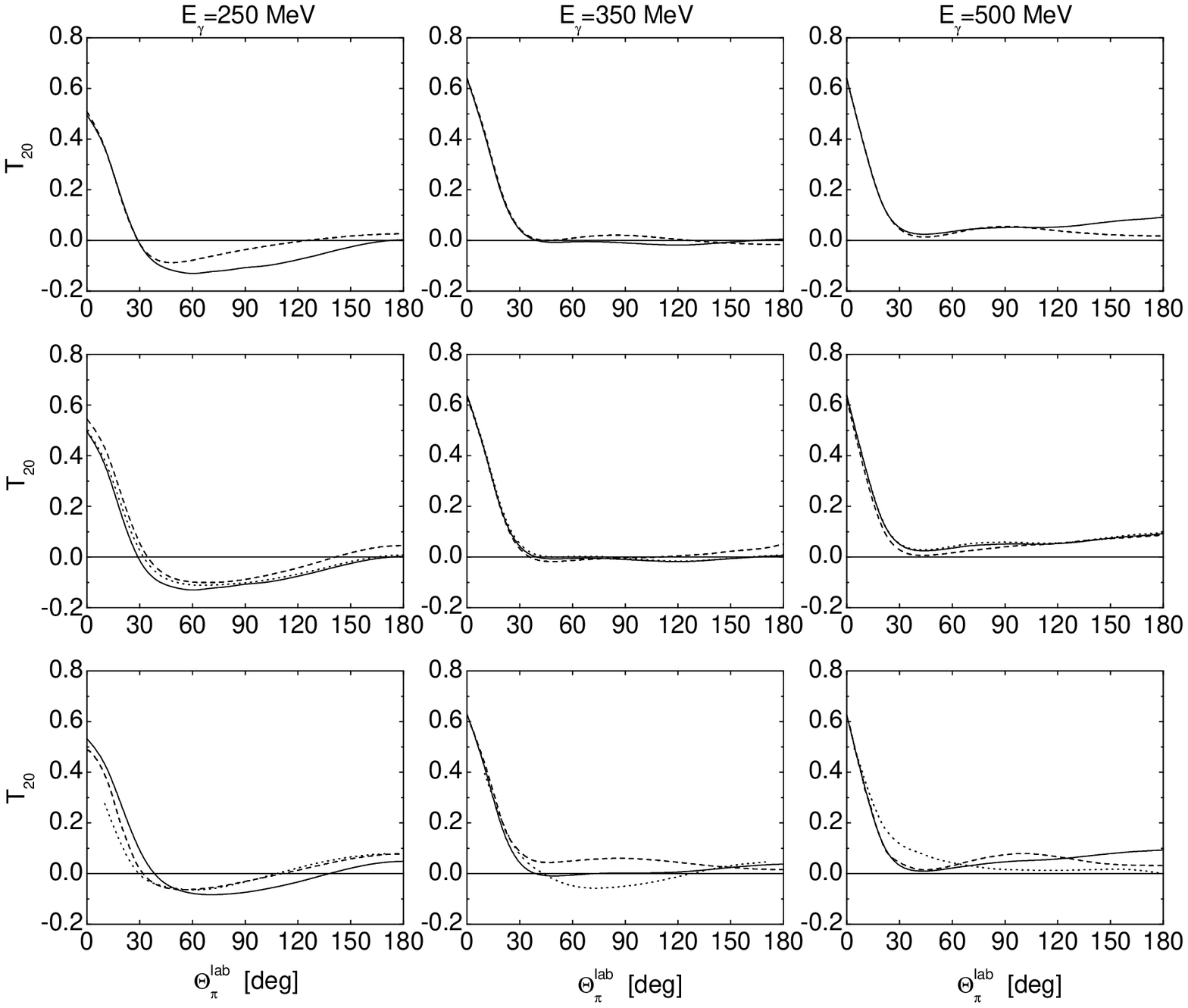}
\caption
{
Target asymmetry $T_{20}$ for the reaction $d(\gamma,\pi^-)pp$.
Notation described in the legend to Fig.~\ref{fig17}.
}
\label{fig19}
\end{figure}

\begin{figure}[hbt]
\includegraphics[width=0.6\textwidth]{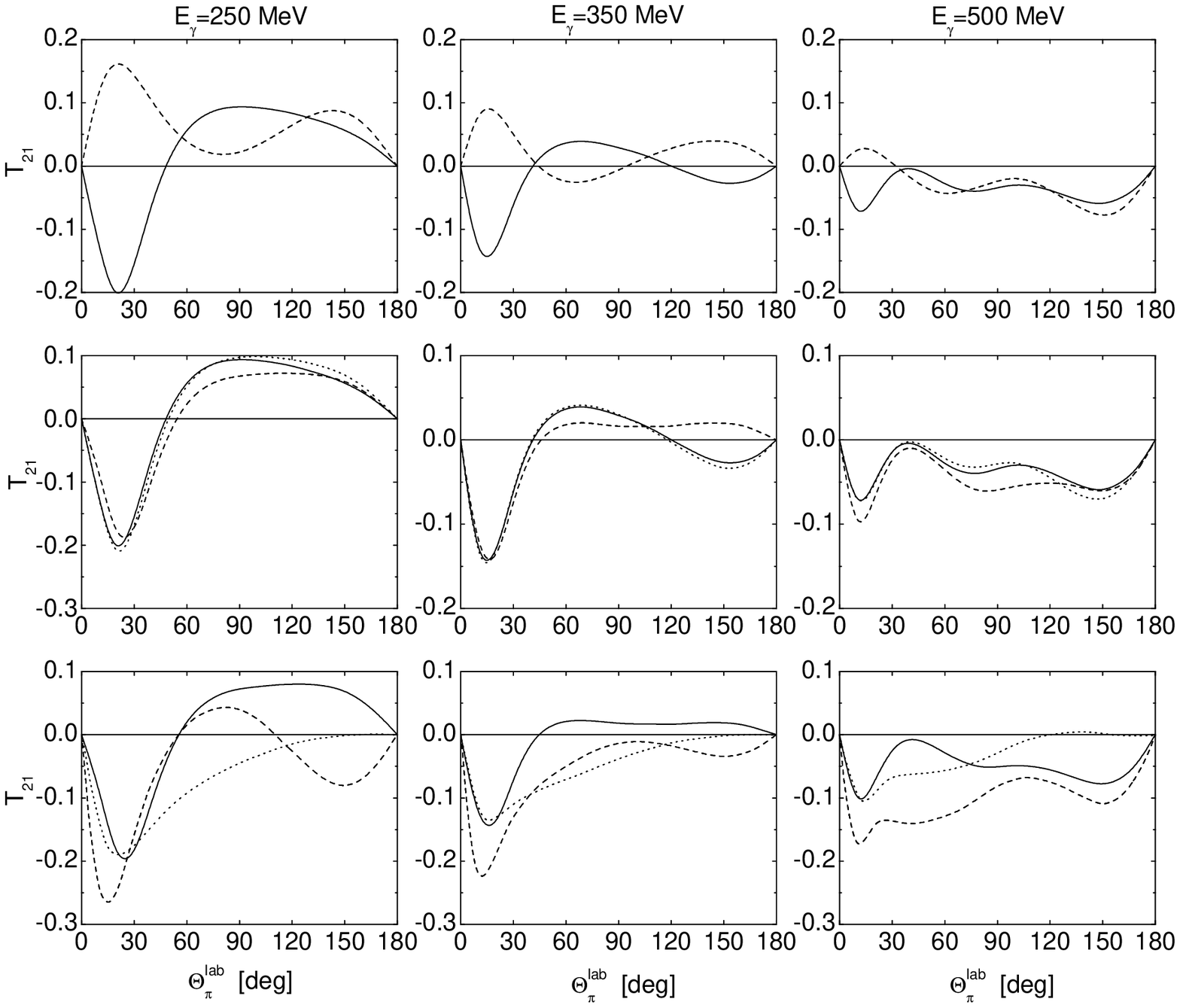}
\caption
{
Target asymmetry $T_{21}$ for the reaction $d(\gamma,\pi^0)np$.
Notation described in the legend to Fig.~\ref{fig12}.
}
\label{fig20}
\end{figure}

\begin{figure}[hbt]
\includegraphics[width=0.6\textwidth]{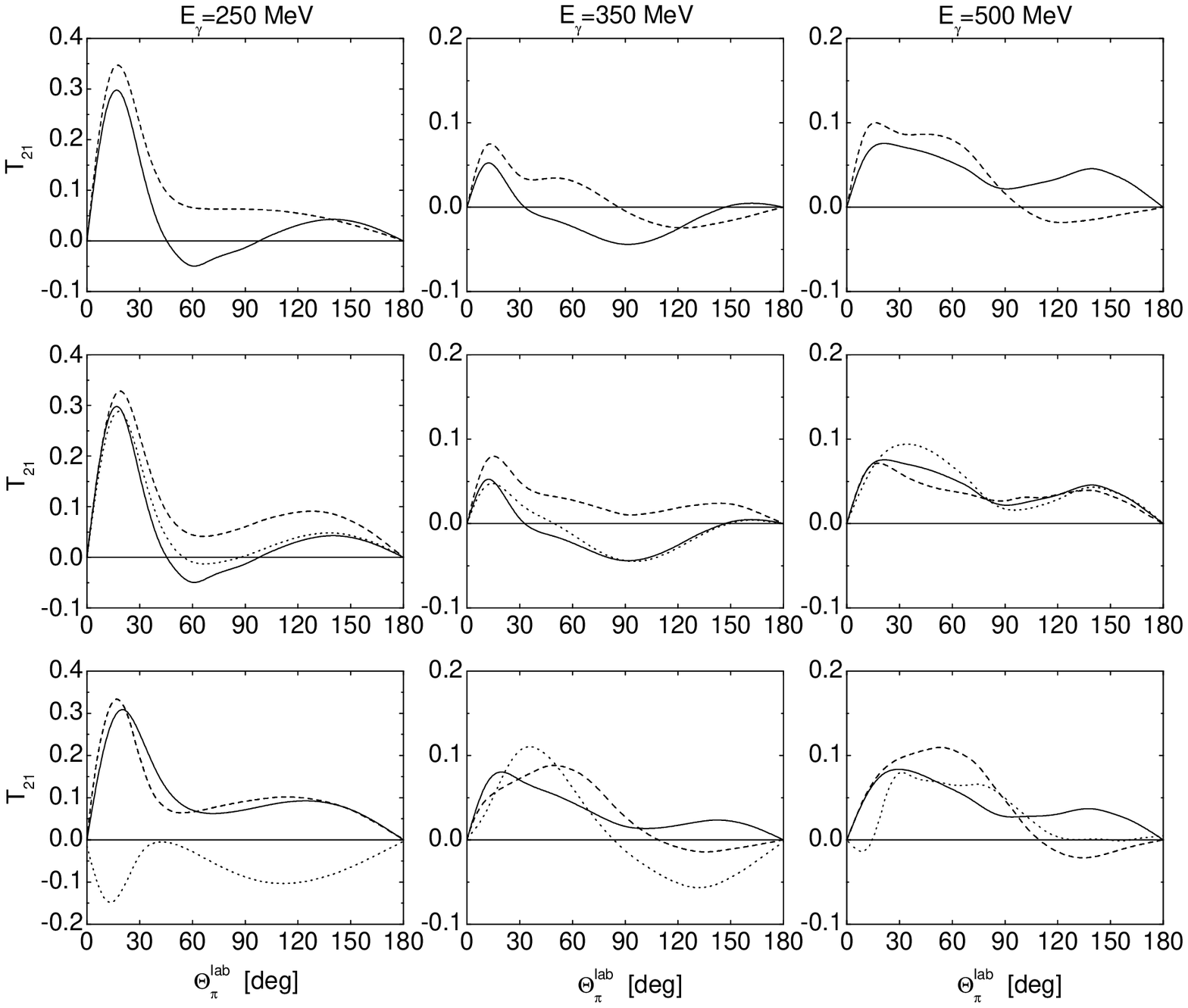}
\caption
{
Target asymmetry $T_{21}$ for the reaction $d(\gamma,\pi^-)pp$.
Notation described in the legend to Fig.~\ref{fig17}.
}
\label{fig21}
\end{figure}

\begin{figure}[hbt]
\includegraphics[width=0.6\textwidth]{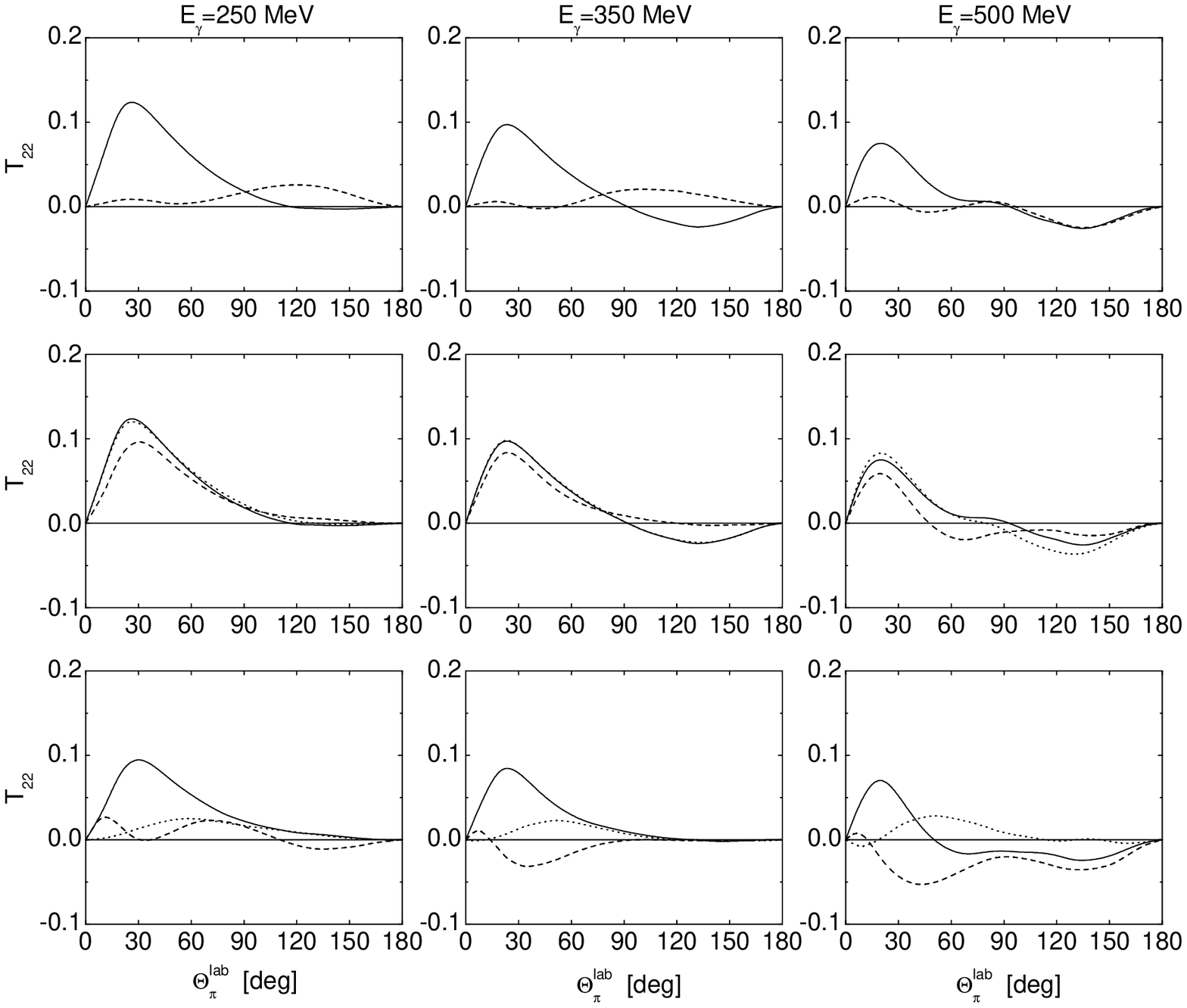}
\caption
{
Target asymmetry $T_{22}$ for the reaction $d(\gamma,\pi^0)np$.
Notation described in the legend to Fig.~\ref{fig12}.
}
\label{fig22}
\end{figure}

\begin{figure}[hbt]
\includegraphics[width=0.6\textwidth]{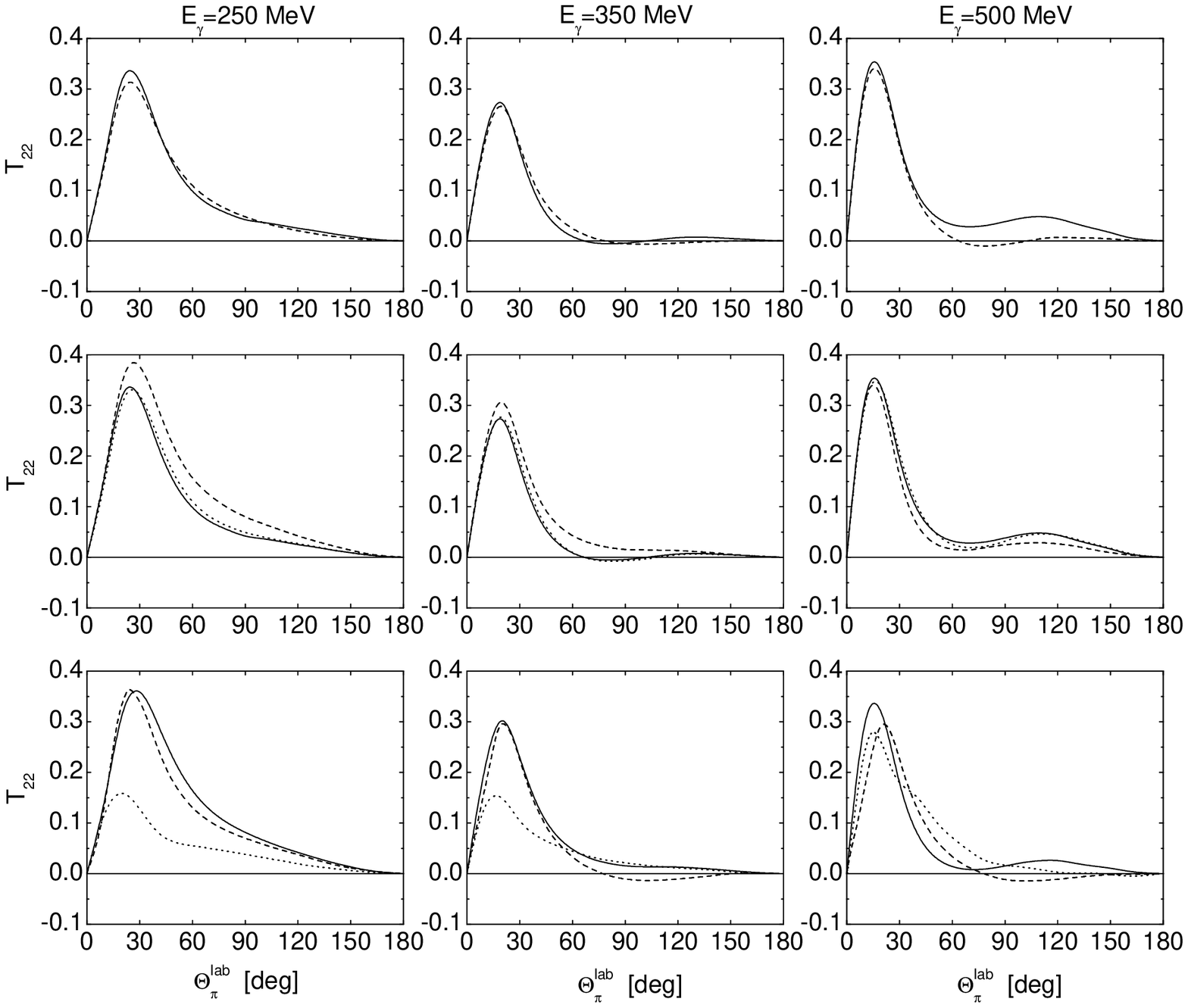}
\caption
{
Target asymmetry $T_{22}$ for the reaction $d(\gamma,\pi^-)pp$.
Notation of the curves as in Fig.~\ref{fig17}.
}
\label{fig23}
\end{figure}

The target asymmetries $T_{IM}$ for $\pi^0$ and $\pi^-$
production are shown in Figs.~\ref{fig16}-\ref{fig23}.
The asymmetries for $\pi^+$ production are very similar to those
in the $\pi^-$ channel and are not discussed below.

In accordance
with results from Ref.~\cite{FixAr05}, we have found
that the form the angular distribution of $T_{11}$
in the $\pi^0$ channel changes
notable with increasing energy. FSI effects are
rather small. The sensitivity of $T_{11}$ to the choice of the photoproduction
operator is quite small at the lowest energy. This is
not the case at higher energies. The vector asymmetry
is sensitive to both the choice of the analysis and the form of the
operator.
A good agreement of our results with those
from Ref.~\cite{FixAr05} is seen. Only at forward angles
$\Theta_\pi \leq 30^\circ$ we have found some deviation.
The predictions from Refs.~\cite{darwish04,darwish05_1}
totally contradict our results both in the form of
the angular distribution and in the absolute size at 250 and 350 MeV
\footnote
{When making comparisons to results from
Refs.~\cite{darwish04,darwish05_1,darwish05_4}
one should keep in mind that
in those works the asymmetry $T_{11}$ is defined
with the opposite sign.}
. For instance, we observe a maximum around
$130^\circ - 140^\circ$ at 250 and 350 MeV with $T_{11}\simeq 0.4$
in the center of the peak,
whereas in Refs.~\cite{darwish04,darwish05_1} the vector asymmetry
is close to zero at these energies for $\Theta_\pi\geq 90^\circ$.
We have no a reasonable explanation for this disagreement.
Only at the highest energy one has a good agreement between
the three calculations.

Figure~\ref{fig17} shows that FSI effects on the vector asymmetry
for $\pi^-$ production are much smaller than for $\pi^0$ production.
An analogous result has been reported in Ref.~\cite{FixAr05}.
As in the case of the $\pi^0$ channel, the vector asymmetry
is practically independent of the photoproduction
operator at the lowest energy. At higher energies we observe
some sensitivity of $T_{11}$ to that operator
especially in the angular region between $30^\circ$ and $120^\circ$.
One can observe very good agreement with the results of
Ref.~\cite{FixAr05} but significant  disagreement
with those of Ref.~\cite{darwish05_4}
is evident. For instance, we do not find a peak near
$30^\circ$ at 250 MeV predicted in Ref.~\cite{darwish05_4}.
In the peak position our value for $T_{11}$ is
by a factor of $\sim 3$ smaller than that in Ref.~\cite{darwish05_4}.

The tensor asymmetry $T_{20}$ for $\pi^0$ production is displayed
in Fig.~\ref{fig18}. It is small in absolute size both in IA and
IA+FSI. The FSI effect is very large at forward angles as it was
previously found also in the case of other observables
in the neutral channel.
The SAID and MAID models give very close predictions for $T_{20}$
but the results are sensitive to a form of a production operator.
The asymmetry $T_{20}$ is the first observable for which we
have found a substantial deviation from results of Ref.~\cite{FixAr05}.
Our resulting asymmetry shows a sharp minimum at forward angles,
whereas a sharp maximum was found in Ref.~\cite{FixAr05}.
The model \cite{darwish05_4} predicts even deeper minimum at forward
angles than that in our calculation.

As is seen in Fig.~\ref{fig19}, the asymmetry $T_{20}$ for the
$\pi^-$ channel is forward peaked in IA. FSI has only a
marginal impact on $T_{20}$. The resulting asymmetry is
very small for $\Theta_\pi\geq 30^\circ$ especially at high
energies. In this kinematic region the results are practically
independent of the choice of the production operator.
The three approaches predict close peak values of $T_{20}$.
Some deviation takes place in the regions where
the asymmetry is small.
Note that recently the tensor target asymmetries in  $\pi^-$
photoproduction on the deuteron have been studied
in Ref.~\cite{loginov}.
Because the authors considered the exclusive reaction,
a comparison of our predictions with the results
from that work is impossible.

The target asymmetry $T_{21}$ for $\pi^0$ production in Fig.~\ref{fig20}
shows drastic FSI influence. The resulting absolute size of $T_{21}$
is small.  It does not exceed 0.1 for $\Theta_\pi\geq 30^\circ$ at
all energies and shows a notable model dependence.
There exist quite significant differences between
the results of our model and those from Refs.~\cite{darwish05_4}
and \cite{FixAr05}.
As is seen in Fig.~\ref{fig21}, practically all the above
conclusions remain to
be valid for $\pi^-$ production too. Only at 250 MeV we observe
a smaller influence of FSI and a good agreement with the results of
Ref.~\cite{FixAr05} is evident at this energy.

Predictions for the target asymmetry $T_{22}$
for $\pi^0$ production are shown in Fig.~\ref{fig22}. This
asymmetry is very small in IA. Its absolute size is less than
0.03 in the kinematic region under
consideration. FSI manifests itself in a pronounced peak
around $20^\circ$ although even in the center of the peak
the asymmetry $T_{22}$ is still small being less than 0.12.
The sensitivity of $T_{22}$ to the production operator varies
with the kinematics. It is quite small at 250 and 350 MeV
but becomes to be notable at 500 MeV. It is seen in Fig.~\ref{fig22}
that the differences to the results of the three approaches
are quite significant.

More pronounced peaks around $20^\circ$ but in IA are seen
in Fig.~\ref{fig23}, where the target asymmetry $T_{22}$
for $\pi^-$ production is displayed. Influence of FSI is
quite small and seen only at the highest energy for
$\Theta_\pi \geq 60^\circ$. One notes a sizable dependence of the
predictions to the production operator at the lowest energy.
It is much smaller at higher energies. We find good
agreement with Ref.~\cite{FixAr05} but significant differences
to Ref.~\cite{darwish05_4} are evident at 250 and 350 MeV.


\section{Summary and outlook}
\label{summary}

In the present work we have studied incoherent pion
photoproduction on the deuteron in the first resonance
region taking into account diagrams with a plane-wave
final state and with $NN$ interaction in the final state.
Particular emphasis has been laid on the discussion of
possible uncertainties introduced into the model by
the elementary operator of pion photoproduction
on the nucleon. We have demonstrated that the use of
different forms of the on-shell operator has
a notable impact on predictions both
for the unpolarized cross section and for polarization
observables, in particular, the beam and target asymmetries.
It is evident that these uncertainties will manifest themselves
in the corresponding variations of the amplitude of pion production
on the neutron extracted from deuteron data. We have not
studied beam-target asymmetries in the present work
leaving their consideration for a subsequent publication.
However, we are confident that analogous uncertainties will
be also seen in these asymmetries.

We have also carried out a detailed comparison of our predictions
for unpolarized cross sections and beam and target asymmetries
with recent results from
Refs.~\cite{darwish04,darwish05_1,darwish05_3,darwish05_4}
and Ref.~\cite{FixAr05}. For most observables we have found
good agreement with results of the latter work. However,
as a rule our predictions are in significant deviation from
those in the former works. Of course, part of the disagreement
can stem from the use of a different elementary production
operator. But in many cases the deviation is too big to have
such an explanation so that the reasons for the deviation are
still to be understood.

Practically for all parametrizations of the photoproduction
operator, our predictions for the unpolarized differential
and total cross sections in the $\pi^0$ channel are
too big in comparison to the available data. The agreement with data
in the $\pi^-$ channel is quite good. The situation with the
description of the preliminary data from the LEGS collaboration
on the beam asymmetry $\Sigma$ is opposite. The agreement
is satisfactory  for $\pi^0$ production but
our predictions for the $\pi^-$ channel clearly
underestimate the experimental values in absolute size.

An important problem to be solved is the significant
disagreement between the theory and experimental
data from Refs.~\cite{armstr72,maccorm96} on the
photoabsorption cross section for deuterium in the vicinity of the
$\Delta$ peak. Even with the allowance for all the uncertainties
considered in the present model, we have found our predictions to
overestimate significantly the measured values.
We suppose that the problem might
be in the data themselves. Using the deuteron values,
the authors of Ref.\ \cite{armstr72} extracted
the total photoabsorption cross section on the neutron
that, near the $\Delta$ resonance energy, is
by $\sim 20\%$ lower than that presently predicted by
the SAID and MAID analyses. New measurements of the deuteron
cross section in the peak region would be of very importance
to clarify the situation.

Much additional work should be also done to improve
the theoretical model. In particular, one should try to take
into account two-loop diagrams. As has already been mentioned
in Refs.~\cite{laget81,kossert04} for the case of exclusive pion
production, there are the kinematic regions where
two-loop diagram with $NN$ rescattering in the intermediate
state can be of importance. However, it is not an easy task
to include that diagram in the model for inclusive
production because numerical calculations become to be
extremely time consuming.
Also, in view of the notable sensitivity of the predictions
for many observables to the choice of the elementary operator,
there is urgent need to develop realistic models which take
into account of off-shell effects in pion photoproduction
on bound nucleons.

\begin{acknowledgments}

Valuable discussions with A.I. Fix and A.I. L'vov are highly  appreciated.
We are very grateful to S.S. Kamalov for providing us with the results
of his calculations on coherent pion photoproduction
on the deuteron and to R. Machleidt for a computer code for the
CD-Bonn potential.  This work was supported by
Belarus RFFR  under contracts F05-168 and F05-303,
by Deutsche Forschungsgemainschaft under
contracts SCHU222 and 436 RUS 113/510, and
by the Russian Foundation for Basic Research grant 05-02-17080.

\end{acknowledgments}

\appendix
\section{An elementary pion photoproduction operator}
\label{operator}

The invariant pion photoproduction amplitude can be written as
\begin{eqnarray}
T_{\gamma N\to \pi N'}
&=&
\bar{u}(p')
\Big[
\sum_{i=1}^{4} A_{i}(s,u,t)\, \Gamma_{i}
\Big]
\,u(p),
\label{TAi}
\end{eqnarray}
where
\beq
s=(k+p )^2=(q+p')^2,\quad
u=(k-p')^2=(q-p )^2,\quad
t=(k-q)^2=(p-p')^2,
\eeq
and
\begin{eqnarray}
\Gamma_{1}
& = & i \gamma_{5}\,\Ge\,\Gk,
\nn
\Gamma_{2}
& = & i \gamma_{5} \left[ q \cdot \epsilon\, (p+p')\cdot k
- q\cdot k \,\,(p+p') \cdot \epsilon \right]\,,
\nn
\Gamma_{3}
& = & i \gamma_{5} \left( q\cdot k \,\Ge
- q \cdot \epsilon \, \Gk \right)\,,
\nn
\Gamma_{4}
& = &   \epsilon_{\mu \nu \rho \sigma} \gamma^{\mu} q^{\nu}
\epsilon^{\rho} k^{\sigma}\,.
\label{TGamma}
\end{eqnarray}
The matrix $\gamma_5$ and antisymmetric tensor
$\epsilon_{\mu \nu \rho \sigma}$
are fixed according to the conditions
\beqn
\gamma_5=+\Big(
\begin{tabular}{cc}
 0 & 1 \\
 1 & 0 \\
\end{tabular}
\Big)\quad \rm{and}\quad
 \epsilon_{0123}=+1.
\eeqn

In the spinor form, the matrix $T$ reads
\beq
\langle m_2|T|\lambda m_1\rangle=
\langle m_2|L+i{\vec\sigma\cdot\vec K}|\lambda m_1\rangle.
\label{LK}
\eeq
Contributions of the amplitudes $A_i$ to the matrix $T$
(\ref{LK}) in an arbitrary frame are as follows
\beqn
L_1&=&NN'\Big[
-\frac {\vec p \cdot\vec S}{E_+}
+\frac {\vec p~'\cdot\vec S}{E_+'}
+\omega \frac { \e \cdot \vec C }{E_+E_+'}
\Big]~A_1,
\nn
\vec K_1&=&NN'\Big[\e \Big(
\omega + \omega \frac { \vec p \cdot \vec p~' }{E_+E_+'}
-\frac {\vec p \cdot\vec k}{E_+}
-\frac {\vec p~'\cdot\vec k}{E_+'}\Big)+
\vec k
\Big(
\frac {\vec p \cdot\e}{E_+} +
\frac {\vec p~'\cdot\e}{E_+'}
\Big)
- \vec p ~\omega \frac {\vec p~' \cdot\e}{E_+E_+'}
- \vec p'~\omega \frac {\vec p  \cdot\e}{E_+E_+'}
\Big]~A_1,
\nn
L_2&=&0,
\nn
\vec K_2&=&2NN'\Big(
\e \cdot \vec p~' \,p  \cdot k-
\e \cdot \vec p  \, p' \cdot k\Big)
\Big(\frac {\vec p}{E_+}-\frac {\vec p~'}{E_+'}\Big)~A_2,
\nn
L_3&=&-NN'\frac 1{E_+E_+'}
\big[
 \vec q \cdot \e \,\vec k \cdot \vec C +
      q \cdot k      \,\e \cdot \vec C
\big]~A_3,
\nn
\vec K_3&=&NN'\Big[
\Big(1-\frac { \vec p \cdot \vec p~' }{E_+E_+'}\Big)
(\e \,q \cdot k + \vec k\, \vec q \cdot \e )+
\nn
&&
\vec p ~\Big(\Big(
-\frac {\omega}{E_+} +\frac {\vec p~' \cdot \vec k}{E_+E_+'}
\Big) \vec q \cdot \e +
\frac {\vec p~' \cdot \e\, q\cdot k }{E_+E_+'}\Big)+
\vec p~' ~\Big(\Big(
-\frac {\omega}{E_+'} +\frac {\vec p \cdot \vec k}{E_+E_+'}
\Big) \vec q \cdot \e +
\frac {\vec p \cdot \e\, q\cdot k }{E_+E_+'}\Big)
\Big]~A_3,
\nn
L_4&=&NN'
\frac 1{E_+E_+'}
\Big[
- \e \cdot \vec C \,\, \omega (E_+ +E_+') +
\vec p  \cdot \vec S \,\,
( E_+'^2 + \vec p \cdot \vec p~')-
\vec p~' \cdot \vec S \,\,
(E_+^2 + \vec p \cdot \vec p~')
\Big]~A_4,
\nn
\vec K_4&=&NN'\Big[
 -\e \Big(
 q^0\Big(
\frac {\vec p \cdot \vec k}{E_+} -
\frac {\vec p~'\cdot \vec k}{E_+'}
     \Big)-
      \omega \Big(
\frac {\vec p \cdot \vec q}{E_+} -
\frac {\vec p~'\cdot \vec q}{E_+'}
     \Big)+
\frac{ \vec p~' \cdot \vec k\,\vec q \cdot \vec p-
 \vec p  \cdot \vec k\,\vec q \cdot \vec p~' }
 {E_+E_+'}
     \Big)+
 \nn
 &&
\vec k
\Big(
 q^0
     \Big(
\frac {\vec p \cdot \e}{E_+} -
\frac {\vec p~'\cdot \e}{E_+'}
     \Big)-
\frac{\e \cdot \vec p \,\vec q \cdot \vec p~' -
\e  \cdot \vec p~'\,\vec q \cdot \vec p }{E_+E_+'}
\Big)
     - \vec q
\Big(
\omega
 \Big(
\frac {\vec p \cdot \e}{E_+} -
\frac {\vec p~'\cdot \e}{E_+'}
 \Big)+
\frac {\vec C \cdot \vec S}{E_+E_+'}
 \Big)
\Big]~A_4,
\label{LiKi}
\eeqn
where $\vec S=\vec k \times \e$,
$\vec C=\vec p \times \vec p~' $,
 $E_\pm=E\pm m$,
$E'_\pm=E'\pm m$, $N=\sqrt{E_+/2m}$, and $N'=\sqrt{E'_+/2m}$.

The photoproduction operator in the c.m. frame has the
well-known form \cite{CGLN}
\beqn
\label{CGLN}
\langle m_2| T_{\gamma N\to \pi N}^\ast|\lambda m_1\rangle=
\frac {4\pi W}m
\langle m_2|
i \vec \sigma\cdot \e_\lambda^{\,\ast} ~F_1 +
\vec \sigma \cdot\hat{\vec q}^{\,\ast}\, \vec \sigma
\cdot (\hat{\vec k}^\ast \times \e_\lambda^{\,\ast})~ F_2 +
i\vec \sigma\cdot \hat{\vec k}^\ast ~
\hat{\vec q}^{\,\ast}\cdot \e_\lambda^{\,\ast} ~F_3 +
i\vec \sigma\cdot \hat{\vec q}^{\,\ast} ~
\hat{\vec q}^{\,\ast}\cdot \vec \e_\lambda^{\,\ast} ~F_4
|m_1\rangle,
\eeqn
where $W=\sqrt{s}$ and the superscript asterisk is
used for the corresponding quantities in the $\gamma N$ c.m. frame.
A comparison of Eq.~(\ref{LiKi}) in the c.m. frame with Eq.~(\ref{CGLN})
gives the following relation between the amplitudes $F_i$ and $A_i$
\begin{eqnarray}
\left(\begin{array}{r}
F_1 \\
\frac {E_+'}{q^\ast} F_2 \\
\frac{1}{q^\ast}  F_3
\\
\frac {1}{E^{\prime}_-} F_4
\end{array}\right) &=& \frac {W_-}{8\pi W}
\sqrt{E^{\prime}_+E_+}
\left(\begin{array}{cccc}
 ~~1 & 0  &\frac{q \cdot k}{W_-}&\frac{W_-^2-q \cdot k}{W_-} \\
 -1 & 0&\frac{q \cdot k}{W_+} &\frac{W_+^2-q \cdot k}{W_+} \\
  0 & ~~W_-  & 1       & -1 \\
0 &  -W_+ & 1 & -1
\end{array}\right)
\left(\begin{array}{c}
 A_{1} \\
 A_{2} \\
 A_{3} \\
 A_{4}
\end{array}\right)\,,
\label{FAamplitude}
\end{eqnarray}
where $W_\pm=W\pm m$.
The inverse relation is as follows
\begin{eqnarray}
\left(\begin{array}{c}
A_1 \\ A_2 \\ A_3 \\A_4
\end{array}\right) &=& \frac {4\pi}{q^\ast\omega^\ast}~
\left(\begin{array}{cccc}
 W_+ &-W_- &-2m \frac{q \cdot k}{W_-}
           &-2m \frac{q \cdot k}{W_+} \\
 0 & 0 & 1 &-1 \\
 1 & 1 & \frac{W_-W_+-q \cdot k}{W_-}
       & \frac{W_-W_+-q \cdot k}{W_+} \\
 1 & 1 & -\frac{q \cdot k}{W_-} & -\frac{q \cdot k}{W_+}
\end{array}\right)
\left(\begin{array}{c}
\sqrt{E_-E^{\prime}_-}\frac{1}{W_-} F_{1} \\
\sqrt{E_+E^{\prime}_+}\frac{1}{W_+} F_{2} \\
\sqrt{\frac{E_+}{E^{\prime}_+}}
\frac{1}{W_+} F_{3} \\
\sqrt{ \frac{E_-}{E^{\prime}_-}}\frac{1}{W_-} F_{4}
\end{array}\right)\,.
\label{AFamplitude}
\end{eqnarray}
Note that the corresponding formula in Ref.~\cite{salam04} contains
missprints for the amplitudes $A_1$ and $A_3$.
One should emphasize that all variables in
Eqs.~(\ref{FAamplitude}) and
(\ref{AFamplitude}) are taken in the c.m. frame, in particular
\beqn
\omega^\ast=\frac {s-m^2}{2\sqrt{s}}=\frac {W_+W_-}{2W}=\sqrt{E_+E_-},
\quad
q^\ast=\frac 1{2W}\sqrt {[W^2-(m+\mu)^2]
                    [W^2-(m-\mu)^2] }=\sqrt{E_+'E_-'}.
\label{omqcm}
\eeqn

Another set of the invariant amplitudes $A_i'$ was used
in Ref.~\cite{FixAr05}
\begin{eqnarray}
T_{\gamma N\to \pi N'}'
&=&
\bar{u}(p')
\Big[
\sum_{i=1}^{4} A_{i}'(s,u,t)\, \Gamma_{i}'
\Big]
\,u(p),
\label{TAi'}
\end{eqnarray}
where
\begin{eqnarray}
\Gamma_1'&=&i \gamma_{5}\,\Ge\,\Gk=\Gamma_1,
\nn
\Gamma_2'&=&2i\gamma_5(\epsilon\cdot p'k\cdot p-
                       \epsilon\cdot p\,k\cdot p')=-\Gamma_2,
\nn
\Gamma_3'&=&i\gamma_5( \Gk\,\epsilon\cdot p -\Ge\,k\cdot p),
\nn
\Gamma_4'&=&i\gamma_5( \Gk\,\epsilon\cdot p'-\Ge\,k\cdot p').
\label{Ai'}
\end{eqnarray}
There are  the following relations between the amplitudes $A_i$ and $A_i'$
\beq
A_1'=A_1-2mA_4,\quad A_2'=-A_2,\quad
A_3'=-A_3-A_4,\quad\ A_4'=A_3-A_4.
\label{Ai'Ai}
\eeq

Using the spinor form of the amplitude $T$ [Eq.~(\ref{LiKi})]
and that for $T'$ given in Ref.~\cite{bennhold},
one can show that the difference between $T$ and $T'$ reads
\beq
T-T'=-A_4\,\frac {NN'}{E_+E_+'}
\langle m_2|\delta \vec p~'\cdot\vec S - \delta '\vec p\cdot\vec S +
i\vec\sigma \cdot[\e \,
(\delta (m\omega+p~'\cdot k)+\delta'(m\omega+p\cdot k))+
\vec k (\delta \vec p~'\cdot \e + \delta' \vec p\cdot \e)]
|\lambda m_1\rangle,
\label{Delta}
\eeq
where $\delta=p^2-m^2$ and  $\delta'=p'\,^2-m^2$.
 Therefore, two representations of the pion production amplitude,
Eqs.~(\ref{TAi}) and (\ref{TAi'}), are equivalent only in the case
of on-shell nucleons when $\delta=\delta'=0$.


\begin{thebibliography}{xx}
\bibitem{legs_np}
Preliminary data of the LEGS collaboration are quoted in Ref.~\cite{FixAr05}.
\end{thebibliography}
\end{document}